\documentclass[nofootinbib,aps,prx,twocolumn,superscriptaddress,showpacs]{revtex4-1}
\usepackage{amsmath,amssymb,graphics,epsfig,epstopdf,color,verbatim,ulem,braket,tabularx}
\usepackage{multirow}
\usepackage[colorlinks,linkcolor=blue,citecolor=blue,urlcolor=blue,bookmarks=false]{hyperref}

\usepackage{listings}
\usepackage{cancel}
\usepackage{mathrsfs}
\usepackage{soul}
\usepackage{color}
\usepackage{url}

\newcommand{\beginsupplement}{%
        \setcounter{table}{0}
        \renewcommand{\thetable}{S\arabic{table}}%
        \setcounter{figure}{0}
        \renewcommand{\thefigure}{S\arabic{figure}}%
}

\makeatletter
\def\maketitle{
\@author@finish
\title@column\titleblock@produce
\suppressfloats[t]}
\makeatother

\begin{document}

\title{Superconductivity enhanced by pair fluctuations between wide and narrow bands}

\author{Changming Yue}
\email{changming.yue@unifr.ch}
\affiliation{Department of Physics, University of Fribourg, 1700 Fribourg, Switzerland}

\author{Hideo Aoki}
\affiliation{Department of Physics, University of Tokyo, Hongo, Tokyo 113-0033, Japan}
\affiliation{Electronics and Photonics Research Institute, National Institute of Advanced Industrial Science and Technology (AIST), Tsukuba 305-8568, Japan}

\author{Philipp Werner}
\email{philipp.werner@unifr.ch}
\affiliation{Department of Physics, University of Fribourg, 1700 Fribourg, Switzerland}

\begin{abstract}
Full or empty narrow bands near the Fermi level are known to
enhance superconductivity by promoting scattering processes and spin fluctuations. Here, we demonstrate that doublon-holon fluctuations in systems with half-filled narrow bands can similarly boost the superconducting $T_c$. We study the half-filled attractive bilayer Hubbard model on the square lattice using dynamical mean-field theory. The band structure of the noninteracting system contains a wide band formed by 
bonding orbitals and a narrow band formed by antibonding orbitals, with bandwidths tunable by the inter-layer hopping. 
The shrinking of the narrow band can lead to a substantial increase in the superconducting order parameter and phase stiffness in the wide band. At the same time, the coupling to the wide band allows the narrow band to remain superconducting -- and to reach the largest order parameter -- in the flat band limit.
We develop an anomalous worm sampling method to study superconductivity in the limit of vanishing effective hopping.
By analyzing the histogram of the local eigenstates, we clarify how the interplay between different interaction terms in the bonding/antibonding basis promotes pair fluctuations and superconductivity. 
\end{abstract}

\maketitle

\newpage

{\it Introduction.}
Superconductivity in strongly-correlated {\it multi-band} systems has attracted much interest since the discovery of iron based superconductors  
\cite{RMP2011,LaOFeP2006,LaOFeAs2008,LiFeAs2008,BaFe2As2,CaFe2As2,KSrFe2As2,NaFeAs,Wang2012FeSe,Liu2012FeSe,He2013FeSe,Tan2013FeSe,Zhang2014FeSe,Zhang2016FeSe}
and also in connection with twisted bilayer graphene \cite{YuanCao2018}.  
Much effort has been devoted to reveal connections between the pairing in systems
with spin, orbital, or nematic degrees of freedom \cite{Kuroki2008,Onari2010,Chubukov2012FeBaseSC,Hoshino2015,ZiXangLi2016,Dumitrescu2016,Huang2017FeseRev,YueFeSe2021}.   
Even the single-orbital square-lattice Hubbard model can be mapped to an effective multi-orbital system \cite{Werner2016,Werner2020}, or 
we can explore non-Bravais lattices \cite{Kuroki2005,Kobayashi2016},  
 which provides novel perspectives and insights into the pairing mechanism. 
Often, the original or effective models exhibit wide and narrow 
bands, which raises the interesting question how the different bandwidths 
cooperate in the superconductivity. 

Recently, it was shown that 
so-called incipient bands \cite{Kuroki2005,Kobayashi2016,Matsumoto2020,Kato2020,Karakuzu2021,Ochi2022,Chen2015}, which are full (empty) bands slightly below (above) the Fermi energy, 
can significantly enhance $T_c$. 
The concept of incipient bands was introduced by Kuroki {\it et al.} \cite{Kuroki2005}  
in a fluctuation exchange (FLEX) 
\cite{Bickers1989} 
study of a Hubbard ladder.
They found that the large number of interband pair-scattering 
channels promotes superconductivity.
Linscheid {\it et al.} \cite{Linscheid2015} argued that the incipient band contributes significantly to the spin-fluctuation pairing
and leads to a high $T_c$ in a two-band system with electron-like and hole-like bands. 
Very recently, Ochi {\it et al.} \cite{Ochi2022} studied a two-band continuum model with incipient narrow empty band with attractive interactions,
and found  
that interband pair-hopping 
induces an effective intraband attraction in each band, enhancing superconductivity.

In the limit where the narrow band becomes flat, the normal-state kinetic energy of the electrons populating this band is quenched. Such (almost) flat bands 
appear in many van der Waals systems, including magic-angle twisted bilayer graphene \cite{YuanCao2018} and its trilayer or double bilayer derivatives 
\cite{Shen2020,YuanCao2020,XLiu2020,XiZhang2021}, and also in twisted bilayer WSe$_2$ \cite{LWang2020} and MoS$_2$ \cite{LedeXian2021}.
This situation has been theoretically suggested to promote superconductivity 
for repulsive interactions \cite{Aoki2020}.   
While most previous works focused on models where either the narrow band 
or wide band 
is empty,
we consider here a situation where {\it all bands are half-filled}. 
Based on the intuition from correlated systems in the normal state,
one might expect that a flat band must be a
Mott insulator (a paired Mott insulator in the case of attractive interactions that we consider here). However, we shall show that, when accompanied by 
a wide band, the flat band can be superconducting (SC) and that the exchange of pairs between the wide and flat bands results in a large SC order parameter in both bands.  

\begin{figure}
\includegraphics[clip,width=3.4in,angle=0]{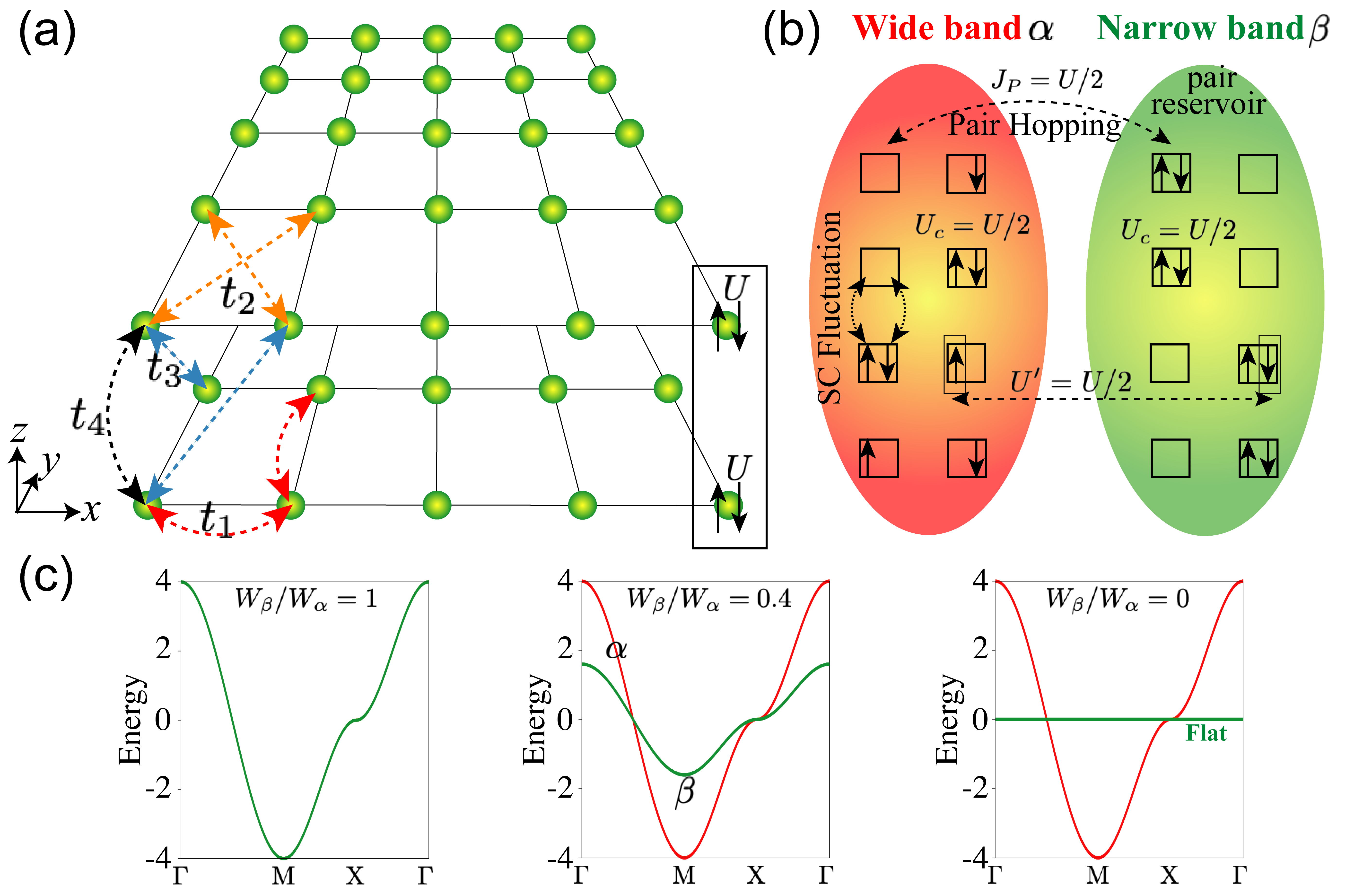}
\caption{
(a) Bilayer square-lattice Hubbard model with sites depicted as green spheres and the two-site unit cell enclosed by a black box. 
$t_1$ (red) is the intra-layer hopping between nearest neighbor sites, 
while $t_2$ (orange) is for second-neighbor sites.  
$t_4$ (black) and $t_3$ (blue) are the inter-layer hoppings between nearest neighbor sites and next-nearest neighbor sites.  
(b) Schematic illustration 
showing the pair fluctuations (scatterings) within the wide band and between the two bands, as well as the relevant interactions. 
Black boxes in (b) represents unit-cells. 
(c) Non-interacting band structures for $W_{\beta}/W_{\alpha}=1$, 0.4, 0.0, respectively. 
}
\label{fig:tb_model}
\end{figure}

{\it Model and Method.} 
We consider the Hubbard model on a bilayer square lattice with an attractive onsite interaction ($U<0$), 
\begin{equation}
H=\sum_{ij,ab}t_{ij}^{ab}c_{i,a\sigma}^{\dagger}c_{j,b\sigma}+U\sum_{ia}n_{i,a\uparrow}n_{i,a\downarrow}.
\end{equation}
Here $a$, $b$ label the layers, and $i$, $j$ the lattice sites, while $\sigma=\,\uparrow,\downarrow$ denotes the spin.
The unit cell of the model contains two sites stacked along the $z$ axis.
The hopping parameters, depicted in Fig.~\ref{fig:tb_model}, 
are the hopping $t_1$ for intra-layer nearest neighbors, $t_2$
for second neighbors, while the inter-layer hoppings are nearest-neighbor 
$t_4$ and second-neighbor $t_3$. 
The non-interacting Hamiltonian $H_{0}^\uparrow({\bf k})=H_{0}^\downarrow({\bf k})$ is diagonal in the bonding-antibonding basis for cell $i$,
$|i,{\alpha\atop\beta}\sigma\rangle=(|i,a\sigma\rangle \pm |i,b\sigma\rangle)/\sqrt{2}$, 
with the bands $\epsilon_{{\alpha\atop\beta},\bf{k}}=\pm t_4+4t_{2}\cos k_{x}\cos k_{y}+2(t_{1}\pm t_{3})(\cos k_{x}+\cos k_{y})$.
If $t_1>0$ and $t_3>0$, $\epsilon_{\alpha}({\bf k})$ 
has a larger bandwidth than $\epsilon_{\beta}({\bf k})$, 
see Fig.~\ref{fig:tb_model}(c).
When $t_2=0$ and $t_3=t_1$, $\epsilon_{\beta}({\bf k})=-t_4$ is 
a flat band.
The hopping $t_4$ determines the energy splitting between the bonding
and anti-bonding bands. Here we set $t_2=t_4=0$ to have 
a particle-hole symmetry.
The band width of each band is $W_{\alpha\atop\beta}=8(t_1\pm t_3)$.
We fix the width of the wide band as $W_{\alpha}=8$, and tune the narrow band width $W_\beta=8(1-2t_3)$ 
by adjusting $t_1$ and $t_3$, and use $W_\alpha/8=1$ as energy unit.

The onsite Hubbard interaction can be transformed, within a unit cell $i$ with two sites, into a two-orbital Hamiltonian,
\begin{align}
&\tilde{H}_{\mathrm{int}}^{i} = U_{\mathrm{c}}\sum_{\alpha}n_{i,\alpha\uparrow}n_{i,\alpha\downarrow}+U^{\prime}\sum_{\alpha\ne \beta}n_{i,\alpha\uparrow}n_{i,\beta\downarrow}\label{eq:Hloc_new}\\
&-J_{P}\sum_{\alpha\ne \beta}c_{i,\alpha\uparrow}^{\dagger}c_{i,\alpha\downarrow}^{\dagger}c_{i,\beta\uparrow}c_{i,\beta\downarrow}
-J_{S}\sum_{\alpha\ne \beta}c_{i,\alpha\uparrow}^{\dagger}c_{i,\alpha\downarrow}c_{i,\beta\downarrow}^{\dagger}c_{i,\beta\uparrow} \, ,\nonumber
\end{align}
with $\alpha$ ($\beta$) the bonding (anti-bonding) orbitals and $U_c=U^\prime=J_P=J_S=U/2$ \cite{Shinaoka2015, Werner2016, Aoki2020}. 
There is no 
inter-orbital same-spin interaction, since $U'-J_S=0$.
The $J_P$ ($J_S$) term describes pair-hopping (spin-flipping) between the bonding and anti-bonding orbitals. 

We solve the interacting lattice model using dynamical mean field theory (DMFT) \cite{Georges1996}, which maps 
the lattice problem to a self-consistently determined Anderson impurity model. 
To solve the two-orbital impurity model in the bonding/anti-bonding basis we employ the hybridization-expansion 
continuous-time quantum Monte Carlo algorithm \cite{Werner2006PRL,Werner2006PRB,Gull2011}. We use four-operator updates to ensure an ergodic sampling in the SC phase \cite{Semon_Ergodicity}.
Furthermore, we developed a normal (anomalous) worm-sampling to measure the normal (anomalous) Green's function for the flat band, since these functions cannot be measured with the conventional technique based on removing (anomalous) 
hybridization lines. 
Details on the anomalous worm algorithms are given in Sec.~5 of the Supplemental Material (SM). 
In the Nambu-formalism, the non-interacting lattice Hamiltonian reads
\begin{equation}
H_{0}=\sum_{\boldsymbol{k}}\!\left[\begin{array}{cc}
\Psi_{\boldsymbol{k},\uparrow}^{\dagger} & \Psi_{-\boldsymbol{k},\downarrow}\end{array}\right]\!\!\left[\begin{array}{cc}
H^{\uparrow}_{0}({\bf k}) & 0\\
0 &-H^{\downarrow}_{0}(-{\bf k})^{T}
\end{array}\right]\!\!\left[\begin{array}{c}
\Psi_{\boldsymbol{k},\uparrow}\\
\Psi_{-\boldsymbol{k},\downarrow}^{\dagger}
\end{array}\right] ,\nonumber
\end{equation}
where we define the Nambu spinors
$[\begin{array}{cc}
\Psi_{\boldsymbol{k},\uparrow}^{\dagger} & \Psi_{-\boldsymbol{k},\downarrow}\end{array}]=[\begin{array}{cccc}
c_{{\bf k},\alpha \uparrow}^{\dagger},& c_{{\bf k},\beta\uparrow}^{\dagger},& c_{-{\bf k},\alpha\downarrow},
& c_{-{\bf k},\beta\downarrow}\end{array}]$.
The interacting lattice Green's function can be expressed as
\begin{equation}
G({\bf k},i\omega_{n})=[i\omega_{n}\mathbb{I}_{4}+\sigma_{3}\otimes \mu\mathbb{I}_{2}-H_{0}({\bf k})-\Sigma^{\mathrm{Nambu}}(i\omega_{n})\mathbb{I}_{4}]^{-1},\nonumber
\label{eq:Gkw}
\end{equation}
where $\Sigma^{\mathrm{Nambu}}$ is the local self-energy from DMFT.  
Unless otherwise mentioned, we set $U=-1$ and $\mu=U/2$ to make the system particle-hole symmetric.

\begin{figure}[t]
\includegraphics[clip,width=3.4in,angle=0]{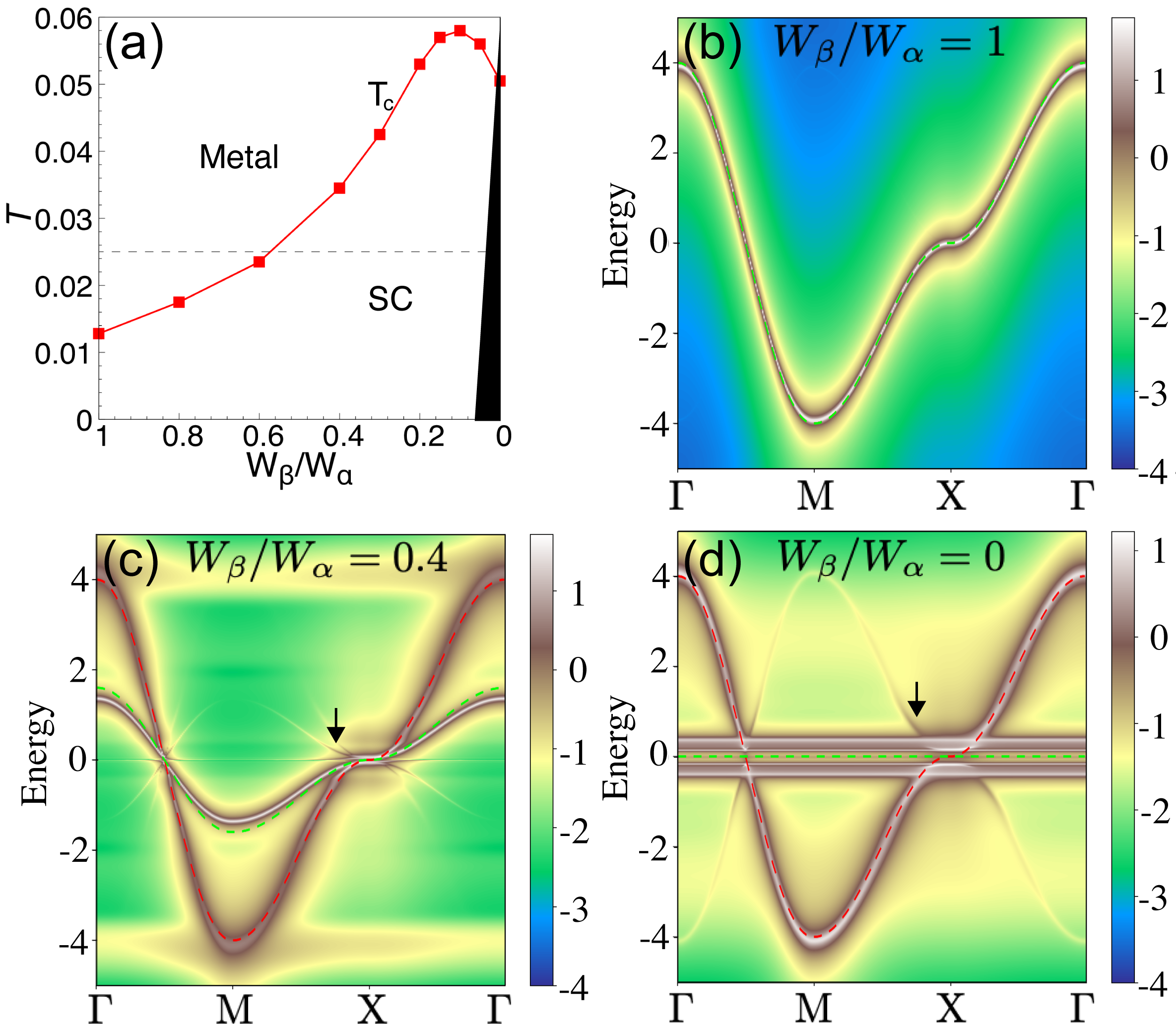}
\caption{
(a) $T_c$ versus the bandwidth ratio.
The black region indicates the Mott phase in the flat band in the normal state. The Mott region extends to $W_\beta/W_\alpha\approx 0.05$ at $T=0.01$. 
(b-d) Momentum-resolved spectral function ${\log}_{10}A(\bf k,\omega)$ for the indicated values of $W_\beta/W_\alpha$ at $T=0.025$ (horizontal dashed line in (a)).
Here, the dashed lines show the non-interacting band structures. 
The black arrows in (c-d) highlight back-bending of the Bogoliubov bands.
}
\label{fig:Tc_Akw}
\end{figure}

{\it Phase diagram and quasi-particle spectra.}  
Figure~\ref{fig:Tc_Akw}(a) presents the DMFT phase diagram in the space of temperature $T$ and bandwidth ratio $W_\beta/W_\alpha$.
Both bands become superconducting simultaneously and we determine $T_c$ by extrapolating the square-root like  
critical behavior of the SC order parameter (see SM Sec.~3).
The red line shows $T_c$ against $W_\beta/W_\alpha$.
For $W_\beta/W_\alpha=1$, we have $T_c\simeq 0.0128$.
In this limit with $t_3=0$ the two layers are decoupled, so that the system 
decomposes into two independent single-band Hubbard models on the square lattice.
As one decreases $W_\beta/W_\alpha$ from 1, 
$T_c$ is seen to increase. 
This can be understood by the decreasing width of the narrow band, 
where $|U|/W_\beta$ increases, i.e.,  
$T_c$ increases with increasing 
electron correlations. 
For $W_\beta/W_\alpha \lesssim 0.4$, $T_c$ markedly increases with decreasing bandwidth ratio and reaches its maximum value of 0.058 (nearly 5 times the $T_c$ at $W_\beta/W_\alpha=1$) around $W_\beta/W_\alpha\simeq 0.1$. Then $T_c$ drops slightly as one further decreases $W_\beta/W_\alpha$ from 0.1 to 0, but it remains
high even when the non-interacting anti-bonding band becomes flat. 
In particular, $T_c$ for the coupled bilayer system with $W_\alpha=8$, $W_\beta=0$ is much higher than for the 
decoupled layers with $W_\alpha=W_\beta=8$.  

We now look at the momentum-resolved spectral function, obtained from the Nambu Green's functions as
$A({\bf{k}},\omega)=-\frac{2}{\pi}\mathrm{Im}[G_{1\uparrow,1\uparrow}+G_{2\uparrow,2\uparrow}]({\bf{k}},\omega)$. 
For the analytic continuation from the Matsubara to the real-frequency axis, we use the auxiliary \cite{Reymbaut2015} maximum entropy \cite{Jarrell1996} method, 
where the real-frequency self-energy $\Sigma(\omega)$ is constructed
from two auxiliary self-energy functions $\Sigma_{\pm}=\Sigma^\mathrm{nor}\pm\Sigma^\mathrm{ano}$ 
which have positive definite spectral weight in the presence of particle-hole symmetry \cite{Gull2013}.
Figure~\ref{fig:Tc_Akw}(b-d) shows the 
spectra for $W_\beta/W_\alpha=1$, $0.4$ and $0$ at $T=0.025$.
For comparison, we overlay the corresponding non-interacting bands. 
The system becomes SC for $W_\beta/W_\alpha\lesssim0.57$ at $T=0.025$, 
as shown in Fig.~\ref{fig:Tc_Akw}(a), and therefore a SC gap opens in both bands in panels (c,d).
There the black arrow 
marks the back-bending of the Bogoliubov bands, which demonstrates particle-hole mixing, a 
fundamental consequence of pair condensation \cite{Schrieffer1964,Campuzano1996}. 
At $W_\beta=0$, the narrow band becomes flat 
but remains superconducting, exhibiting a gap opening. 
The gap size of the $\beta$ band is the same as for the $\alpha$ band 
(see SM Sec.~2 for a detailed analysis)
and substantially smaller than $U$, which shows that this is a SC gap and not a Mott gap.  
There are however two additional flat features with an energy splitting of $\approx 0.8U$, which can be associated with the upper and lower Hubbard bands, as suggested by the comparison to the  Hubbard-I spectrum for the normal state shown in SM Sec.~2.

\begin{figure}[t]
\includegraphics[clip,width=3.4in,angle=0]{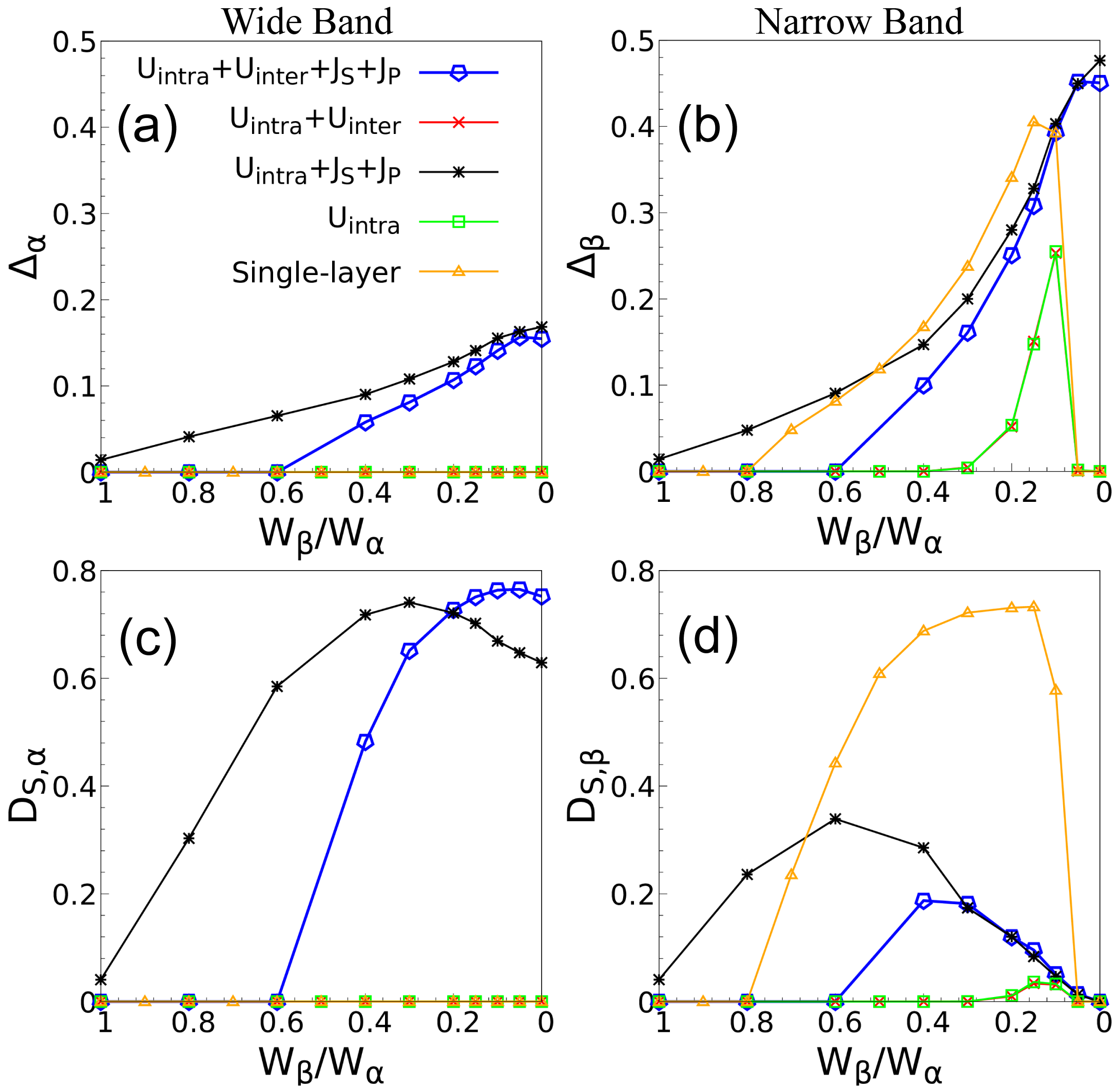}
\caption{
(a,b) SC order parameter $\Delta$ and (c,d) superfluid stiffness $D_S$ (in units of $e^2/\hbar^2$) in the two bands 
against $W_\beta/W_\alpha$ at $T=0.025$. 
Panels (a,c) are for the wide band ($\alpha$) and panels (b,d) for the narrow band ($\beta$). 
Green symbols: results when only intra-orbital interactions are 
considered; 
red: for intra-orbital plus inter-orbital density-density interactions; 
black: for intra-orbital interactions plus spin-flip and pair-hopping terms; 
blue: for the full model.  
Orange symbols in (b,d) [(a,c)] : results for a single-band model with varying bandwidth $W=W_\beta$ [fixed bandwidth $W=W_\alpha=8$].
}
\label{fig:delta_stiff}
\end{figure}

{\it Order parameter and phase stiffness.} 
The phase stiffness $D_{S}$ measures the rigidity of the SC state against phase twisting. We calculate
$D_{S}$ in the framework of linear response and in the long-wave-length limit, following Refs.~\onlinecite{YueFeSe2021,Simard2019,colemanbook} as
$D_{S,xx}=D^{\mathrm{par}}_{S,xx}+D^{\mathrm{dia}}_{S,xx}$
with $D^{\mathrm{par}}_{S,xx}=\frac{e^{2}T}{\hbar^{2} VN}\sum_{{\bf k},i\omega_{n}}\mathrm{Tr} G({\bf k},i\omega_n)(\sigma_0 \otimes\lambda_{{\bf k}}^{x})G({\bf k},i\omega_n)(\sigma_0\otimes\lambda_{{\bf k}}^{x})$ 
and $D^{\mathrm{dia}}_{S,xx}=\frac{e^{2}T}{\hbar^{2} VN}\sum_{{\bf k},i\omega_{n}}\mathrm{Tr}G({\bf k},i\omega_n)e^{i\omega_{n}0^{+}}(\sigma_3\otimes\lambda_{{\bf k}}^{xx})$,
where $\lambda_{{\bf k}}^{x} \equiv \partial_{{\bf k}_{x}} H_0({\bf k})$, and $\lambda_{{\bf k}}^{x x} \equiv \partial^2_{{\bf k}_{x}} H_0({\bf k})$.
A mesh of 395$\times$395 ${\bf k}$-points is used to calculate the stiffness.
The orbital-resolved order parameters $\Delta_\alpha=\langle c_{\alpha\uparrow} c_{\alpha\downarrow} \rangle$ and $\Delta_\beta=\langle c_{\beta\uparrow} c_{\beta\downarrow} \rangle$, and corresponding stiffnesses  $D^{\alpha}_{S,xx}$ and $D^{\beta}_{S,xx}$
($D^\alpha_{S,xx}+D^\beta_{S,xx}=D_{S,xx}$)
are plotted against $W_\beta/W_\alpha$ in Fig.~\ref{fig:delta_stiff} by the blue lines. 
Panels (a,c) show the results for the wide band and panels (b,d) those for the narrow band. We set $T=0.025$, so that the model becomes SC for 
$W_\beta/W_\alpha \lesssim 0.57$. The order parameter and stiffness in the wide band increase with decreasing $W_\beta$ and reach 
respective maxima in or near the flat-band limit $W_\beta=0$. This shows that the stronger correlations in the narrow band 
and the enhanced interband pairing interactions boost superconductivity in the wide band. 
Note that a single-band model with bandwidth 8 and $U=-1$ would not be superconducting at this temperature (orange curves in panels (a,c)). 

In the narrow band, while the order parameter shows a stronger increase reaching its maximum near $W_\beta=0$, the stiffness exhibits a much less pronounced increase than in the wide band, followed by a decrease as the narrow band enters into the strong-correlation regime. 
Remarkably, the narrow band does {\it not} become a paired Mott insulator for small $W_\beta$ unlike in the single-band model (orange curves in panels (b,d)), 
see also the spectra in SM Sec. 4. 
This shows that 
the superconductivity in the narrow band is supported by the interactions with the wide band in the strong-correlation regime.

\begin{figure*}[t]
\includegraphics[clip,width=0.8\paperwidth,angle=0]{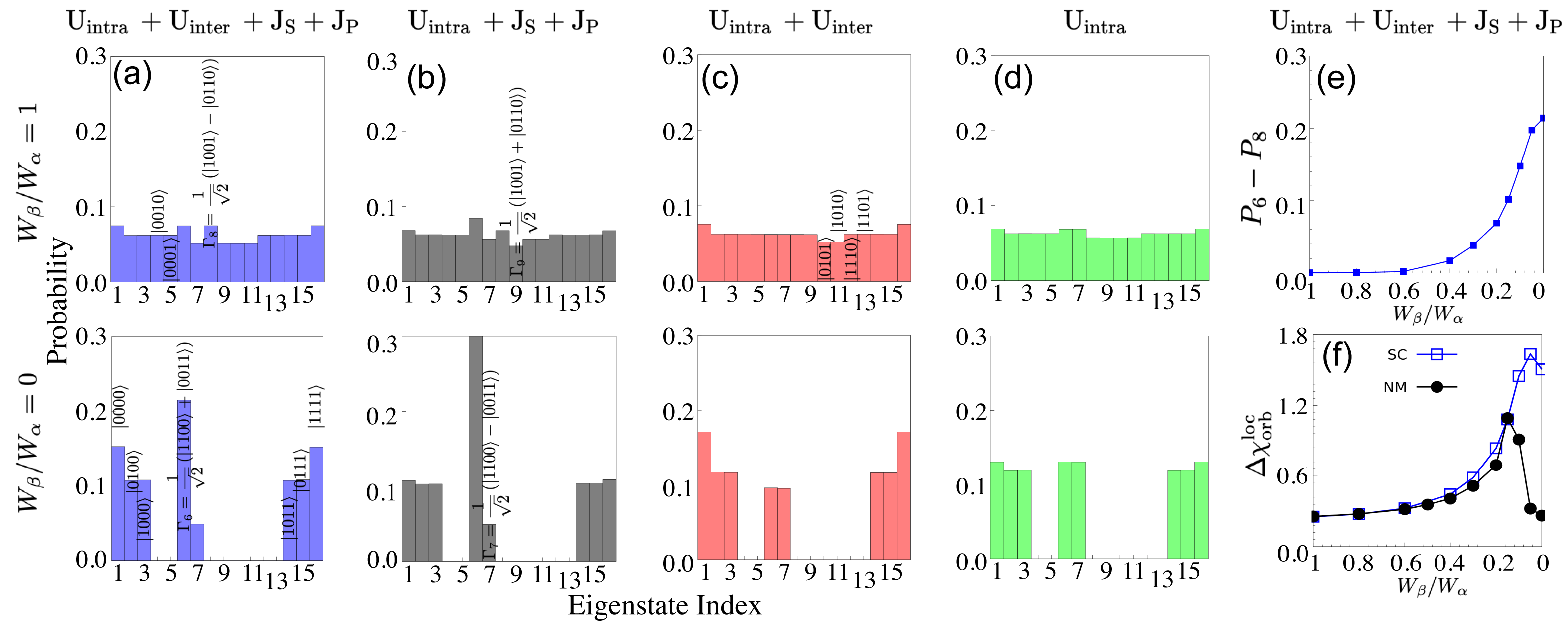}
\caption{ 
(a-d) DMFT histograms of atomic eigenstates 
for models with different interaction terms.
Results are shown for the model with
(a) all the interaction terms, 
(b) the $U_c$, $J_S$, $J_P$ terms, 
(c) the $U_c$, $U^\prime$ terms, and
(d) the $U_c$ term only. 
The top (bottom) row is for $W_\beta/W_\alpha=1$ (0) in (a-d).
(e) Difference in probabilities $P_{\Gamma_6}-P_{\Gamma_8}$ as a function of $W_\beta/W_\alpha$ in the full model. 
(f) Dynamic contribution to the local orbital susceptibility for the full model in the SC and normal metal (NM) phase.  
The temperature is $T=0.025$. 
}
\label{fig:prob}
\end{figure*}
To analyze the mechanism behind the enhancement of superconductivity in the narrow band, let us resolve the effects of the different interaction terms in the effective two-orbital Hamiltonian \eqref{eq:Hloc_new} by turning them on term by term. The green lines in Fig.~\ref{fig:delta_stiff} show the results obtained when we only retain the intra-orbital interaction $U_c$, i.e., for a system without any coupling between the bonding and anti-bonding orbitals. In this case, the only relevant quantities are the ratios $U_c/W_\alpha$ and $U_c/W_\beta$.  
Since we decrease $W_\beta$ at fixed $W_\alpha$, we see the behavior expected for the single-band attractive Hubbard model: the order parameter in the wide band remains constant, while it increases in the narrow band, up to the Mott transition point at $W_\beta/W_\alpha\simeq 0.1$
(see spectra of the $U_c$ model in SM Sec.~4).  
When we add the inter-orbital interactions $U'$ to the intra-orbital interactions $U_c$ we obtain similar results 
as shown by the red lines in Fig.~\ref{fig:delta_stiff}, which overlap with the green lines
 (the almost negligible effect of $U'$ is because of the small value of $U=-1$).  

If instead we consider $U_c$ and the pair-hopping and spin-flip terms (black lines in Fig.~\ref{fig:delta_stiff}), the results are remarkably different. The order parameters 
in both bands are now larger than for the full model, especially for $W_\beta/W_\alpha$ near $1$, and they increase monotonically with decreasing $W_\beta$. Also the stiffness is strongly enhanced for $W_\beta/W_\alpha \gtrsim 0.4$. Since we are considering here intra-orbital pairing, it is natural to assume that the pair-hopping (rather than spin-flip) term is the relevant player in the observed enhancement of superconductivity. 

To further analyze the interplay between the interaction terms, we look at the probability weights of the 
16 eigenstates of $\tilde{H}_{\mathrm{int}}^{i}$ (Eq.~\eqref{eq:Hloc_new}) \cite{SM}, measured with DMFT. Panels (a-d) in Fig.~\ref{fig:prob} show them for $W_\beta/W_\alpha=1$ (top) and $W_\beta/W_\alpha=0$ (bottom), for the four types of interactions with the same color code as in Fig.~\ref{fig:delta_stiff}.  
We label the eigenstates $\Gamma$ using a binary code of the occupation status per spin-orbital
$|n_{\alpha\uparrow}n_{\alpha\downarrow} n_{\beta\uparrow} n_{\beta\downarrow}\rangle$  
as indicated in the figure.
In panel (a) we see that, for $W_\beta/W_\alpha=1$, 
the eigenstate 
$\Gamma_6\equiv\frac{1}{\sqrt{2}}(|1100\rangle+|0011\rangle)$
of the pair-hopping term $H_{P}=-J_{P}\sum_{\alpha\ne \beta}c_{i,\alpha\uparrow}^{\dagger}c_{i,\alpha\downarrow}^{\dagger}c_{i,\beta\uparrow}c_{i,\beta\downarrow}$ 
is as important as the eigenstate 
$\Gamma_8=\frac{1}{\sqrt{2}}(|1001\rangle-|0110\rangle)$ of the
spin-flip term $H_{S}=-J_{S}\sum_{\alpha\ne \beta}c_{i,\alpha\uparrow}^{\dagger}c_{i,\alpha\downarrow}c_{i,\beta\downarrow}^{\dagger}c_{i,\beta\uparrow}$,
while for $W_\beta/W_\alpha=0$, $\Gamma_6$, with a combination of 
inter-band pair-hopped states, clearly dominates. 
In the model without the $U'$ term (panel (b)), $\Gamma_6$ is already more relevant than $\Gamma_8$ at $W_\beta/W_\alpha=1$ 
and it completely dominates for $W_\beta/W_\alpha=0$. 

The pair hopping term boosts superconductivity, as seen from $\Delta$ in Fig.~\ref{fig:delta_stiff}, 
as long as the pairs have a large phase stiffness (are sufficiently delocalized).
A too dominant $\Gamma_6$ state, as in the case of $W_\beta/W_\alpha\approx 0$ in the model without $U'$, weakens the superfluid stiffness (Fig.~\ref{fig:delta_stiff}(c)). The suppression of $\Delta$ and $D_S$ in the full model with $W_\beta/W_\alpha=1$, compared to the model without $U'$, can be explained from the setting $U_c=U'=J_S=J_P$. The density-density interaction is the same for intra-orbital and inter-orbital opposite-spin pairs, so that both the pair-hopping and spin-flipping terms are active and stabilize the states $\Gamma_6$ and $\Gamma_8$, respectively. $\Gamma_8$ 
however favors inter-orbital pairing and suppresses intra-orbital pairing, 
which explains the smaller order parameter and lower $T_c$ of the full model with $W_\beta/W_\alpha=1$. For $W_\beta/W_\alpha<1$ the symmetry between the bonding and antibonding orbitals is broken and the intra-orbital correlations in the narrow band start to dominate the inter-orbital correlations. 
This leads to a strongly correlated metal with a high probability of doublons and holons in the narrow band of this attractive-$U$ system, 
and suppresses the $\Gamma_8$ states. 
The result is the strong increase in $\Delta$ seen in Fig.~\ref{fig:delta_stiff}(a,b) (blue line).
Meanwhile, the presence of the $U'$ interaction prevents too strong a dominance of the $\Gamma_6$ state 
by favoring the full ($\Gamma_{16}$) and empty ($\Gamma_1$) states. 
Hence the full model with pair-hopping and $U'$ favors, for small enough $W_\beta$, a state which supports pair fluctuations and exhibits a large stiffness (blue line in Fig.~\ref{fig:delta_stiff}(c)).
We can think of the flat band as a reservoir of pairs, which are injected into the wide band via pair-hopping processes, thus boosting superconductivity in the wide band. 
At the same time, the pair-hopping enables a kind of proximity effect \cite{footnote}, 
which allows the narrow band to remain superconducting even in the flat-band limit.  
To support the relevance of this mechanism, we plot in Fig.~\ref{fig:prob}(e) the difference $P_6-P_8$ between the probabilities of $\Gamma_6$ and $\Gamma_8$. 
The strong upturn around $W_\beta/W_\alpha \approx 0.4$ is qualitatively similar to the increase seen in 
$\Delta_{\alpha,\beta}$.

A second factor that plays a role in the pairing is the enhancement of the attractive interactions through local moment fluctuations. For a weak enough bare interaction, this effect can be captured by calculating an effective screened interaction which takes into account bubble diagrams, as demonstrated in several works \cite{Inaba2012,Hoshino2015,Steiner2016,Hoshino2017,Yue2021}. Within the random phase approximation, the effective static interactions are given as 
$\tilde{J}_{P,S}=(U/2)/[1-\frac{U}{2}(\chi_{1212}^{\uparrow\uparrow}+\chi_{2121}^{\uparrow\uparrow})\frac{U}{2}(\chi_{1212}^{\downarrow\downarrow}+\chi_{2121}^{\downarrow\downarrow})]$ 
and $\tilde{U}_c,\tilde U^\prime=(U/2)/[1-\frac{U}{2}(\chi_{1111}^{\uparrow\uparrow}+\chi_{2222}^{\uparrow\uparrow})\frac{U}{2}(\chi_{1111}^{\downarrow\downarrow}+\chi_{2222}^{\downarrow\downarrow})]$, 
where $\chi_{pqst}^{\sigma\sigma}(\Omega=0)=-T\sum_{m}G_{ps}^\sigma(i\omega_{m})G_{tq}^\sigma(i\omega_{m})$.
In the weak-coupling limit, all the effective interactions are enhanced by the third-order term in $U$, 
and this effect is augmented in the narrow-band regime if $\chi$ itself 
increases with decreasing $W_\beta$. In the density sector, $\chi$ is related to the orbital susceptibility $\chi_\text{orb}$ \cite{Yue2021}. Since the orbital moments in our effective two-orbital model can freeze in the strong-correlation regime \cite{Steiner2016}, we replace $\chi_\text{loc}^\text{orb}(\Omega=0)$ with the fluctuation contribution to the DMFT orbital correlation function, $\varDelta \chi_\text{loc}^\text{orb}=\int_0^\beta d\tau \chi_\text{loc}^\text{orb}(\tau)-\beta \chi_\text{loc}^\text{orb}(\beta/2)$.  
As shown in Fig.~\ref{fig:prob}(f), $\varDelta \chi_\text{loc}^\text{orb}$ 
in the normal phase (circles) grows with decreasing 
$W_\beta/W_\alpha$, and reaches its maximum 
around $W_\beta/W_\alpha=0.16$ before the narrow band becomes Mott insulating
and the local orbital moments freeze. 
The orbital-frozen metal state has a large entropy \cite{Yue2020,Yue2021}, which is released if the system goes into a SC phase.  
In the SC phase (empty squares in panel (f)), $\varDelta \chi_\text{loc}^\text{orb}$ continues to increase sharply with decreasing $W_\beta/W_\alpha$  
reaching a maximum closer to the flat-band limit. 
The feedback of the enhanced orbital fluctuations on the effective attraction contributes to the boosting of $T_c$ in the narrow- and flat-band regimes. The dip in $\varDelta\chi^\text{loc}_\text{orb}$ near $W_\beta=0$ may explain the similar dip seen in $T_c$ (red curve in Fig.~\ref{fig:Tc_Akw}(a)).

So far we have employed the bonding/anti-bonding basis, but we can readily translate the SC order parameters back to the original site basis. Since $c_{{\alpha\atop\beta}\uparrow}c_{{\alpha\atop\beta}\downarrow}=\frac{1}{2}(c_{a\uparrow}\pm c_{b\uparrow})(c_{a\downarrow}\pm c_{b\downarrow})=\frac{1}{2}(c_{a\uparrow}c_{a\downarrow}+c_{b\uparrow}c_{b\downarrow}\pm c_{a\uparrow}c_{b\downarrow}\pm c_{b\uparrow}c_{a\downarrow})$, with $a$, $b$ labeling the layers, 
and $\Delta_\beta>\Delta_\alpha$ for $W_\beta/W_\alpha<1$, one generically finds that
$\Delta_{aa}=\Delta_{bb}=\frac{1}{2}(\Delta_\alpha+\Delta_\beta)$
and $\Delta_{ab}=\Delta_{ba}=\frac{1}{2}(\Delta_\alpha-\Delta_\beta)\ne0$.  
The system with $W_\beta=0$ exhibits both local pairing with amplitude $\frac{1}{2}(\Delta_\alpha+\Delta_\beta)$ and  
inter-layer spin-singlet pairing $\langle c_{a\uparrow}c_{b\downarrow}-c_{a\downarrow}c_{b\uparrow}\rangle=\frac{1}{2}(\Delta_\alpha-\Delta_\beta)$. 
At $W_\beta/W_\alpha=1$, we have instead $\Delta_{aa}=\Delta_{bb}=\Delta_\alpha=\Delta_\beta$ with $\Delta_{ab}=\Delta_{ba}=0$, and thus only intra-site pairing, 
as expected for decoupled layers. 

{\it Conclusions.} We have demonstrated significant enhancements of superconductivity associated with the interplay between wide and narrow bands. In a half-filled and particle-hole symmetric system with attractive interactions, the strong correlations in the narrow or flat band favor doublons and holons, whose injection into and interaction with the wide band boosts the superfluid stiffness and $T_c$. By a kind of proximity effect (pair-hopping term in the bonding/antibonding basis), the superconductivity in the wide band supports the superconductivity in the narrow band even in the flat-band limit. The pairing is additionally boosted by local orbital fluctuations which effectively enhance the attractive interactions. 

Our results are not related to topological effects \cite{Torma2015,Randeira2021}, since the bandstructure of the bilayer system is non-topological and the wide band features no gap. 
The findings are also qualitatively different from the previous results related to incipient bands \cite{Kuroki2005,Chen2015,Kobayashi2016,Matsumoto2020,Kato2020,Karakuzu2021,Ochi2022}, since these works considered the effects of full or empty narrow bands in repulsive models and found that the half-filled situation does not favor superconductivity \cite{Matsumoto2020}.  
Our bandstructure and the bonding/antibonding transformation used to study the interaction effects is related to previous analyses of the square lattice Hubbard model \cite{Werner2016} and diamond chain \cite{Kobayashi2016}. It will be interesting to extend the results of this study to repulsive systems by investigating the role of narrow bands as a reservoir or seed of local moments, and to clarify the effects on superconductivity induced by local moment fluctuations.

{\it Acknowledgements. ---}
The calculations were performed on the Beo05 cluster at the University of Fribourg, using a code based on iQIST \cite{HUANG2015140,iqist}. C.Y. and P.W. acknowledge support from SNSF Grant No. 200021-196966. 
H.A. thanks CREST (Core Research for Evolutional
Science and Technology project from Japan Science and Technology
Agency; Grant Number JPMJCR18T4).


%
%


\clearpage

\beginsupplement
\title{Supplementary Information\\ Superconductivity enhanced by pair fluctuations between wide and narrow bands}
\maketitle

\begin{widetext}
\begin{flushleft}

\section*{SM1. Eigenstates and eigenvalues of $\tilde H_\text{int}$ for different combinations of interaction terms}

Table \ref{tab:eigen} lists all the eigenvectors and eigenvalues for the local
Hamiltonians with different interaction terms discussed in the main text. We use 4 bits with values
0, 1 to represent the Fock states 
$|(b\!\uparrow) (b\!\downarrow) (a\!\uparrow) (a\!\downarrow\rangle)$
where the first (last) two bits 
give the occupations in the bonding (antibonding) orbital with spin $\uparrow$ and $\downarrow$,
respectively.

The eigenvalues for the different models and the parameters considered in the main text are plotted in Fig.~\ref{fig:eval}.  

\begin{table*}[htp]
\centering
\caption{Eigenvectors and eigenvalues of the local Hamiltonian with different interaction terms.}
\label{tab:eigen}
\medskip
\begin{tabular}{|c|c|c|c|c|c|}
\hline 
\multirow{3}{*}{Index} & \multirow{2}{*}{Eigenvector} & \multicolumn{4}{c|}{Interaction Form, Eigenvalue}\tabularnewline
\cline{3-6} \cline{4-6} \cline{5-6} \cline{6-6} 
 
 &  & $U_{c}=U^{\prime}=J_{S}=J_{P}=\frac{U}{2}$ & $U_{c}=J_{S}=J_{P}=\frac{U}{2}$ & $U_{c}=U^{\prime}=\frac{U}{2}$ & $U_{c}=\frac{U}{2}$        \tabularnewline
 &  & $$		  			       & $U^{\prime}=0$            & $J_{S}=J_{P}=0$ 		    & $U^{\prime}=J_{S}=J_{P}=0$ \tabularnewline
\hline 
\hline 
1     & $|0\rangle$				 		     & 0                          & 0                     & 0 	                      & 0            \tabularnewline
\hline 
2,3   & $|1000\rangle,|0100\rangle$              		     & $-\mu$                     & $-\mu$                & $-\mu$                    & $-\mu$       \tabularnewline
\hline 
4,5   & $|0010\rangle,|0001\rangle$ 		 		     & $-\mu$                     & $-\mu$                & $-\mu$                    & $-\mu$       \tabularnewline
\hline 
6,7   & $\frac{1}{\sqrt{2}}\left(|1100\rangle\pm|0011\rangle\right)$ & $U_{c}\pm J_{P}-2\mu$      & $U_{c}\pm J_{P}-2\mu$ & $U_{c}-2\mu$              & $U_{c}-2\mu$ \tabularnewline
\hline 
8,9   & $\frac{1}{\sqrt{2}}\left(|1001\rangle\mp|0110\rangle\right)$ & $U^{\prime}\pm J_{S}-2\mu$ & $\pm J_{S}-2\mu$      & $U^{\prime}-2\mu$         & $-2\mu$      \tabularnewline
\hline 
10,11 & $|0101\rangle,|1010\rangle$ 				     & $-2\mu$                    & $-2\mu$               & $-2\mu$ 	              & $-2\mu$      \tabularnewline
\hline 
12,13 & $|1110\rangle,|1101\rangle$   				     & $U_{c}+U^{\prime}-3\mu$    & $U_{c}-3\mu$          & $U_{c}+U^{\prime}-3\mu$   & $U_{c}-3\mu$ \tabularnewline
\hline 
14,15 & $|1011\rangle,|0111\rangle$   				     & $U_{c}+U^{\prime}-3\mu$    & $U_{c}-3\mu$          & $U_{c}+U^{\prime}-3\mu$   & $U_{c}-3\mu$ \tabularnewline
\hline 
$16$  & $|1111\rangle$ 					             & $2U_{c}+2U^{\prime}-4\mu$  & $2U_{c}-4\mu$         & $2U_{c}+2U^{\prime}-4\mu$ & $2U_{c}-4\mu$\tabularnewline
\hline 
\end{tabular}
\end{table*}

\begin{figure*}
\includegraphics[clip,width=6.8in,angle=0]{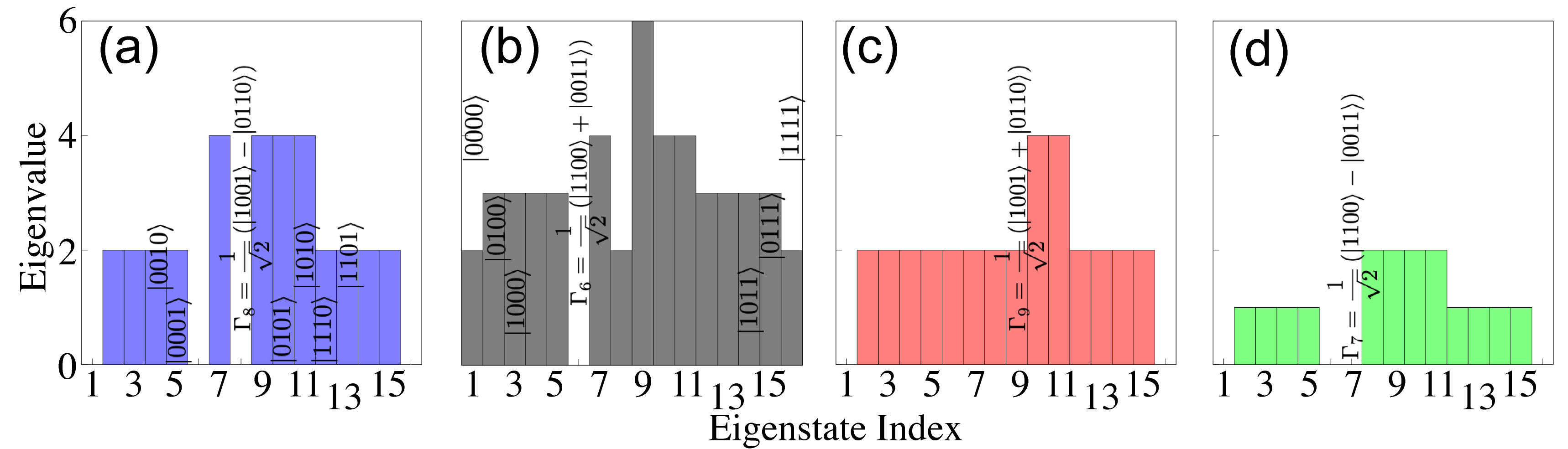}
\caption{
Distribution of eigenvalues for all states listed in Tab.~\ref{tab:eigen}. 
The minimum value for each model is shifted to 0 for ease of comparison.
(a) Result for all interaction terms ($\#6$ column in Tab.~\ref{tab:eigen}). 
(b) Result for the $U_c$, $J_S$ and $J_P$ terms ($\#5$ column in Tab.~\ref{tab:eigen}).
(c) Result for the $U_c$ and $U^\prime$ terms ($\#4$ column in Tab.~\ref{tab:eigen}).
(d) Result for the $U_c$ term only ($\#6$ column in Tab.~\ref{tab:eigen}). 
The parameters are $U=-4$, $\mu=-2$, $t_4=0$.
}
\label{fig:eval}
\end{figure*}

\section*{SM2. SC gap versus Mott Gap}
There is a gap opening both in the spectrum of a Mott insulator and of a superconductor.
As discussed in the main text, the spectra in 
Fig.~2(d) correspond to a SC state. 
In Fig.~\ref{fig:Akw_HubI_vs_DMFTSC}(a) we show the spectrum $A(\bf{k},\omega)$ in the normal phase, 
obtained with the Hubbard-I approximation, 
which employs an atomic-limit self-energy. 
In this case, the flat band is Mott insulating with a gap size of $U$, whereas the wide band shows only a tiny gap.
However, the spectral function in the SC phase obtained with DMFT is distinctly different, as shown in 
Fig.~\ref{fig:Akw_HubI_vs_DMFTSC}(b) (reproduced from Fig.~2(d)).
The gap size of the flat band in the SC phase ($\sim0.4U$) is significantly smaller than the Mott gap $U$ in the atomic limit, 
and the same as the gap of the wide band. 
Hence, we can identify the gap in  Fig.~2(d) as a SC gap rather than a Mott gap.
In the DMFT spectra one can however also notice two dispersionless features at higher energies (with a splitting $\sim 0.8U$),  
as clearly seen in the local spectrum shown in Fig.~\ref{fig:Akw_HubI_vs_DMFTSC}(c), which we interpret as the Hubbard bands. 

\begin{figure*}
\includegraphics[clip,width=6.8in,angle=0]{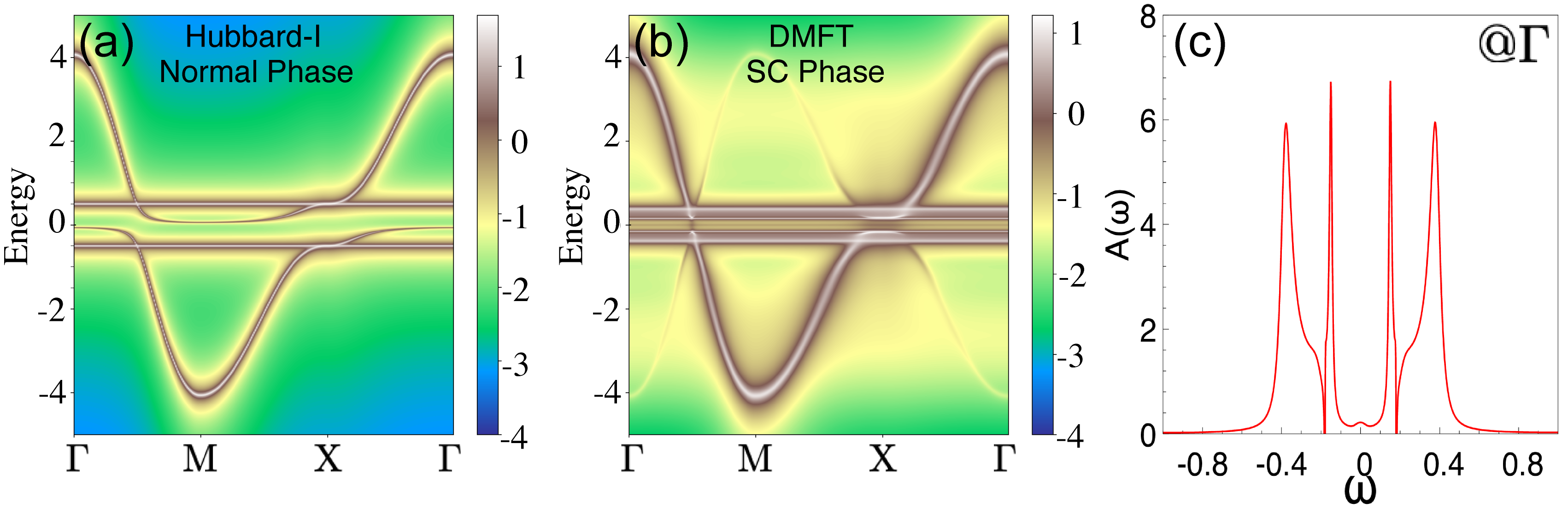}
\caption{
The momentum-resolved spectral function $A(\bf{k},\omega)$ for $W_\alpha=8$ and $W_\beta=0$. 
(a) $A(\bf{k},\omega)$ calculated using the Hubbard-I approximation in the normal phase.
(b) $A(\bf{k},\omega)$ and (c)  $A(\bf{k}=\Gamma,\omega)$ calculated by DMFT in the SC phase.
}
\label{fig:Akw_HubI_vs_DMFTSC}
\end{figure*}
In Fig.~\ref{fig:Aw_Aanow_T0p025_gapsize}, we plot both the normal spectral function $A(\omega)$ and anomalous spectral function $A^\mathrm{ano}(\omega)$
obtained by using the auxiliary \cite{Reymbaut2015} maximum entropy \cite{Jarrell1996} method.
We link the the peak positions in $A(\omega)$ and $A^\mathrm{ano}(\omega)$  by the vertical dashed lines, 
to show that they match. 
These results demonstrate that the SC gap size of the $\beta$ band is the same as that of the $\alpha$ band, 
even though the order parameters are different ($\Delta_\beta>\Delta_\alpha$).
\begin{figure*}
\includegraphics[clip,width=6.8in,angle=0]{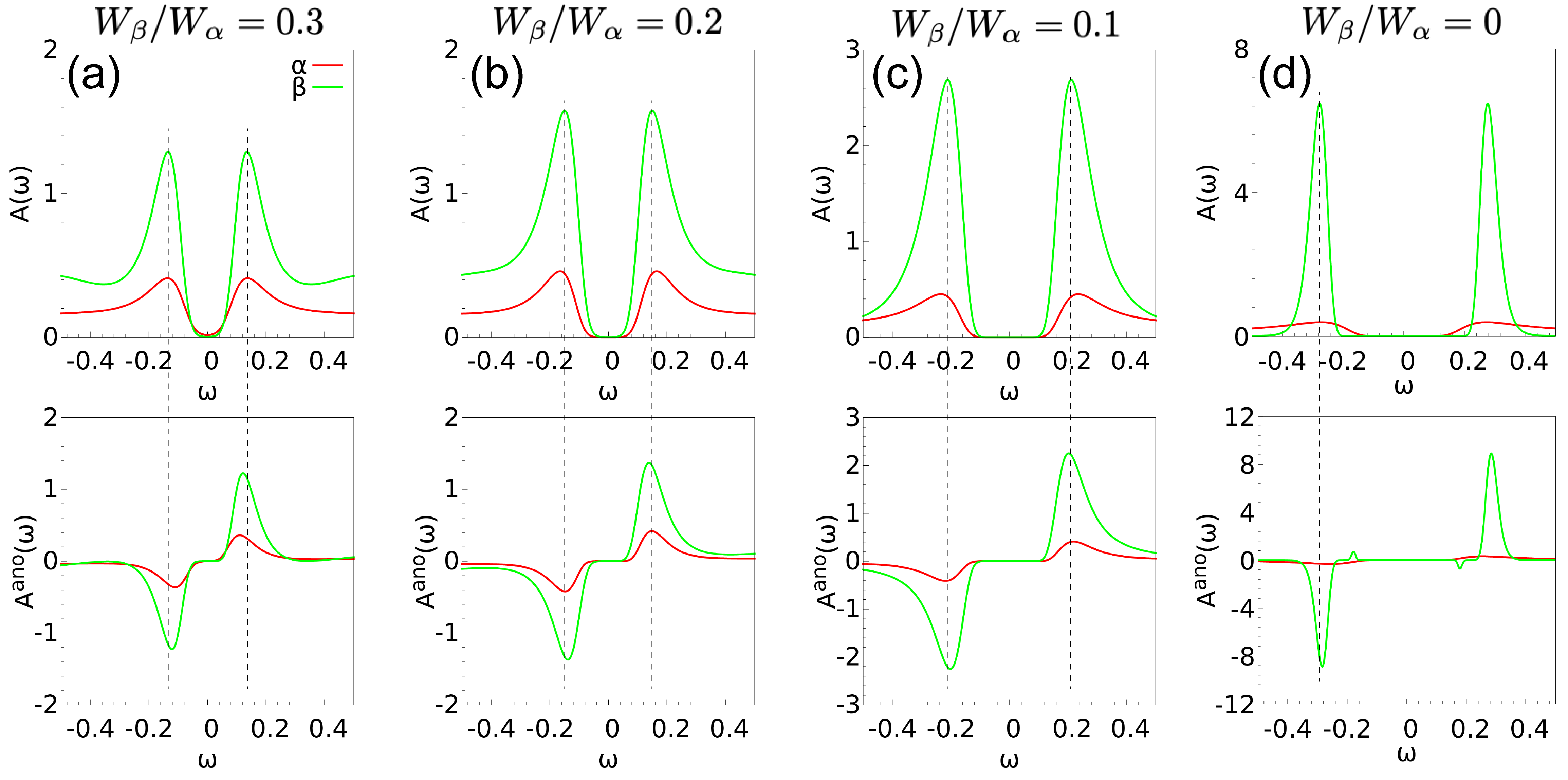}
\caption{
Normal spectral function $A(\omega)$ and anomalous spectral function $A^\mathrm{ano}(\omega)$
for the indicated band-width ratio (red line: bonding orbital $\alpha$, green line: anti-bonding orbital $\beta$).  
The vertical dashed lines mark the peak positions. 
Here, $T=0.025$.
}
\label{fig:Aw_Aanow_T0p025_gapsize}
\end{figure*}

\section*{SM3. SC order parameters and determination of $T_c$}

Fig.~\ref{fig:Delta_Tx}(a) [(c)] shows the superconducting order parameter $\Delta_\alpha$ for the wide
band and $\Delta_\beta$ for the narrow band as a function of $T$ at $W_\beta/W_\alpha=1$ [=0]. As one
decreases $T$, both bands become simultaneously superconducting  
($\Delta_\alpha$ and $\Delta_\beta$ become nonzero at the same $T$), which means that 
$T_c$ is the same for both bands.
 Since the transition from the normal phase to the SC phase
is expected to be second order, we plot $\Delta^2$ as a function $T$ in panels (b,d). 
By extrapolating $\Delta^2$ by a function linear in $T$, we determine $T_c$ as the intersection with the $T$-axis, as shown
by the black dashed lines in panels (b,d). 

\begin{figure*}
\includegraphics[clip,width=7.2in,angle=0]{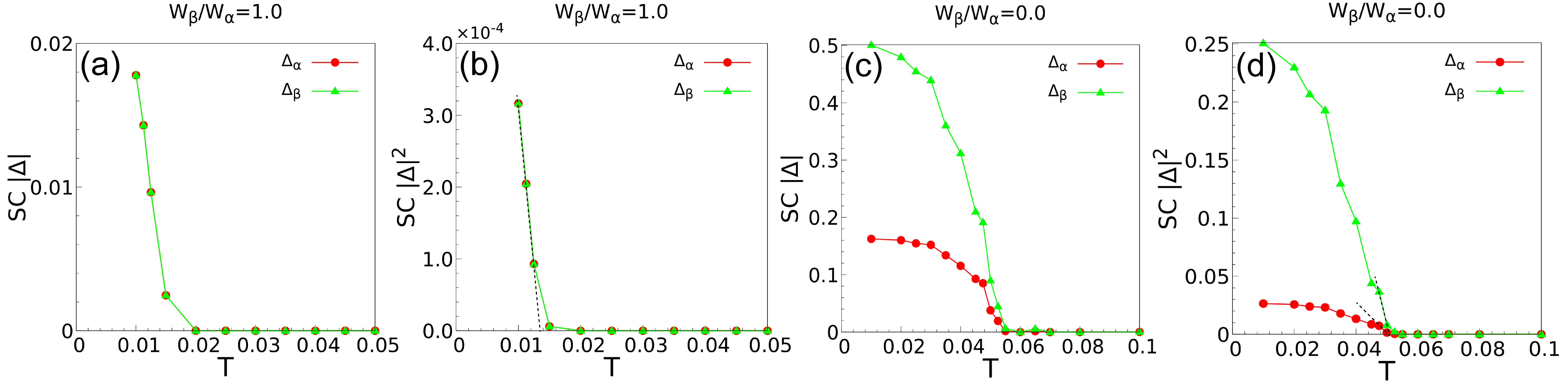}
\caption{
The superconducting order parameter $\Delta$ as a function of $T$.
(a) $\Delta$ and (b) $\Delta^2$ for $W_\beta/W_\alpha=1$.
(c) $\Delta$ and (d) $\Delta^2$ for $W_\beta/W_\alpha=0$.
}
\label{fig:Delta_Tx}
\end{figure*}

\section*{SM4. Spectra in the bilayer Hubbard model and single band Hubbard model}

In Fig.~\ref{fig:spectra_2L_vs_1L}, we compare the spectra of the normal phase for the bilayer Hubbard model and the single-band Hubbard model at small 
$W_\beta/W_\alpha$ (the band width in the single-band Hubbard model is $W\equiv W_\beta$; $W_\alpha=8$ is a constant). At $W_\beta/W_\alpha$=0.1 (panel (a)), the bilayer Hubbard model is in a good metallic state with a 
sharp peak in $A_\beta(\omega)$ (green line), while the single-band model is already in the Mott insulator phase (blue line).
At $W_\beta/W_\alpha$=0.05 (panel (b)), the bilayer Hubbard model becomes a bad metal, with the anti-bonding orbital pseudo-gapped.
The single-band Hubbard model, with a clear gap and well-separated Hubbard bands, becomes more insulating.

\begin{figure*}
\includegraphics[clip,width=4.0in,angle=0]{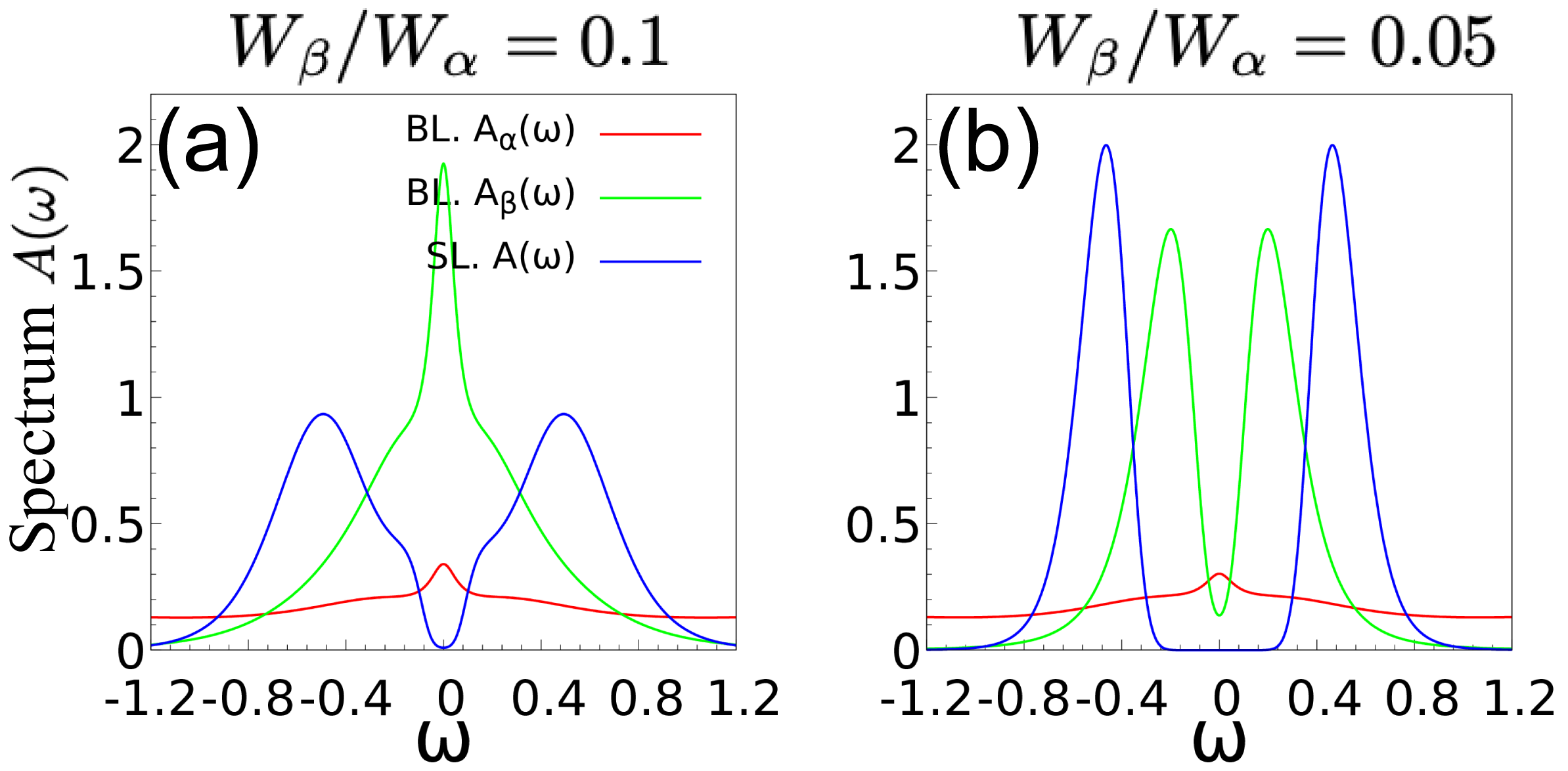}
\caption{
Spectral functions near the Fermi energy in the normal phase of the bilayer (BL.) Hubbard model 
(red line: bonding orbital $\alpha$, green line: anti-bonding orbital $\beta$)
and the single (SL.) band Hubbard model (blue line) for the indicated
band width ratio. Here, $T=0.025$. 
}
\label{fig:spectra_2L_vs_1L}
\end{figure*}

\section*{SM5. Normal and Anomalous Worm Sampling}

Before presenting a detailed discussion of the `anomalous' worm sampling, we briefly
review the conventional worm sampling method for 
hybridization-expansion continuous-time quantum Monte-Carlo algorithm (CT-HYB). This method was
introduced by Gunacker {\it et al.} \cite{Gunacker2015,Gunacker2016}  to measure the one- and two-particle normal
Green's functions with high precision.
It is particularly useful in the atomic limit where
the hybridization function vanishes.
In this case, the standard measurement procedure for the normal Green's function, 
which is based on removing hybridization lines from configuration space $\mathcal{C}_{Z}$ diagrams
of the partition function $Z$, cannot be applied. 
Worm sampling overcomes this limitation by extending the configuration space to 
$\mathcal{C}=\mathcal{C}_{z}\oplus \mathcal{C}_{G^{(1)}}$,
where $\mathcal{C}_{G^{(1)}}$ is the configuration space of the modifed ``partition function" $\mathcal{Z}_{G^{(1)}}$ which is obtained by integrating over all
degrees of freedom of the normal Green's function $G_{\alpha_{1}\alpha_{2}}(\tau_{1},\tau_{2})=-\langle\mathcal{T}_{\tau}c_{\alpha_{1}}(\tau_{1})c_{\alpha_{2}}^{\dagger}(\tau_{2})\rangle$, 
\begin{equation}
Z_{G^{(1)}}=\sum_{\alpha_{1}\alpha_{2}}\int d\tilde{\tau}_{1}d\tilde{\tau}_{2}G_{\alpha_{1}\alpha_{2}}(\tilde{\tau}_{1},\tilde{\tau}_{2}).
\end{equation}
This allows to define the extended partition function $W=Z+\eta_{G^{(1)}}\mathcal{Z}_{G^{(1)}}$, where $\eta_{G^{(1)}}$ is a weighing factor. The difference
between a configuration in $\mathcal{C}_{Z}$ and a configuration
in $\mathcal{C}_{G^{(1)}}$ is that the latter has no hybridization lines
attached to the operators $c_{\alpha_{1}}(\tau_{1})$ and $c_{\alpha_{2}}^{\dagger}(\tau_{2})$,
which are called worm operators. We will refer to the Monte Carlo sampling
in the \emph{extended} configuration space $W$ as (normal) worm sampling.
The insertion and removal of operators are the two basic updates necessary
for an ergodic sampling. Worm replacement/shift updates can further be used
to reduce the auto-correlation time.

\begin{figure*}
\includegraphics[clip,width=4.8in,angle=0]{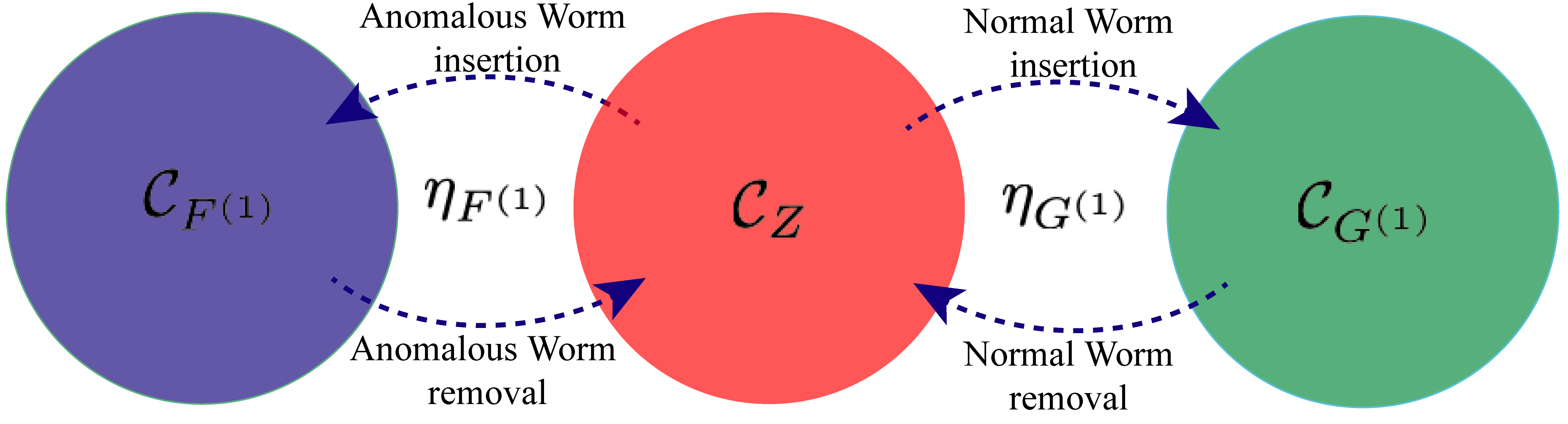}
\caption{
Schematic illustration of the worm sampling in the extended configuration space, which 
includes the partition function space $\mathcal{C}_Z$ (red), the anomalous worm space $\mathcal{C}_{F^{(1)}}$ (blue)
and the normal worm space $\mathcal{C}_{G^{(1)}}$ (green). The weights of the worm space configurations are rescaled by the corresponding weighing factors $\eta_{F^{(1)}}$ and $\eta_{G^{(1)}}$. 
}
\label{fig:worm}
\end{figure*}

In the `anomalous worm sampling', we measure 
$F_{\alpha_{1}\alpha_{2}}(\tau_{1},\tau_{2})=-\langle\mathcal{T}_{\tau}c_{\alpha_{1}}(\tau_{1})c_{\alpha_{2}}(\tau_{2})\rangle$,
the one-particle anomalous Green's function,
using a worm algorithm. We extend the configuration space to include
not only the normal worm space but also the anomalous worm space 
\begin{equation}
\mathcal{C}=\mathcal{C}_{Z}\oplus\mathcal{C}_{G^{(1)}}\oplus\mathcal{C}_{F^{(1)}},
\end{equation}
where we also define a modified ``partition function" $Z_{{F^{(1)}}}$
associated with the configuration space $\mathcal{C}_{F^{(1)}}$ by 
\begin{equation}
Z_{{F^{(1)}}}=\sum_{\alpha_{1}\alpha_{2}}\int d\tilde{\tau}_{1}d\tilde{\tau}_{2}F_{\alpha_{1}\alpha_{2}}(\tilde{\tau}_{1},\tilde{\tau}_{2}).
\end{equation}
The partition function of the extended configuration space reads 
\begin{equation}
W=Z+\eta_{G^{(1)}}Z_{{G^{(1)}}}+\eta_{F^{(1)}}Z_{{F^{(1)}}}.
\end{equation}
As illustrated in Fig.~\ref{fig:worm}, we implement Monte Carlo updates
between the partition function space $\mathcal{C}_Z$ and one of the two worm spaces
by worm operator insertion/removal updates, although direct transitions between the
two worm spaces would also be possible in principle. The updates within each
subspace depend on its structure. Generally, simple pair insertion/removal updates are sufficient for ergodicity.
In practice, however, one finds that worm replacement and shift updates \cite{Gunacker2015}
help to reduce the auto-correlation time. Additional
updates may be necessary if there is a symmetry breaking. For example,
S$\acute{\text{e}}$mon {\it et al.} \cite{Semon2014} showed that is is necessary to perform four-operator
updates in the $d$-wave superconducting state of a single-band repulsive Hubbard model.

In the following, we will show that also in the present attractive-$U$ two-orbital Hubbard system, four-operators updates are necessary
in all subspaces in the SC phase. Furthermore, additional updates
are needed to sample $Z_{{F^{(1)}}}$ due to the imbalanced
number of creation and annihilation operators for each spin flavor.

\subsection*{CT-HYB in the Nambu formalism}

Let us first recall the CT-HYB algorithm in a Nambu formulation appropriate for
intra-orbital spin-singlet pairing.

\subsubsection*{Hamiltonian}

In DMFT, the correlated lattice model in the
SC phase is mapped to an Anderson impurity model with a self-consistently
determined SC bath 
\begin{equation}
H_{\mathrm{AIM}}=H_{\mathrm{loc}}+H_{\mathrm{bath}}^{\mathrm{SC}}+H_{\mathrm{hyb}}.
\end{equation}
We consider the generalized Kanamori local Hamiltonian 
\begin{align}
H_{\mathrm{loc}} & =\sum_{j\sigma}E_{j}\hat{n}_{j\sigma}+U_{c}\sum_{j=1}^{M}\hat{n}_{j\uparrow}\hat{n}_{j\downarrow}+U^{\prime}\sum_{j\ne j^{\prime}}\hat{n}_{j\uparrow}\hat{n}_{j^{\prime}\downarrow}+U^{\prime\prime}\sum_{j>j^{\prime},\sigma}\hat{n}_{j\sigma}\hat{n}_{j^{\prime}\sigma}\nonumber\\
 & -J_{S}\sum_{i,j\ne j^{\prime}}c_{j\uparrow}^{\dagger}c_{j\downarrow}c_{j\downarrow}^{\dagger}c_{j\uparrow}+J_{P}\sum_{j\ne j^{\prime}}c_{j\uparrow}^{\dagger}c_{j\downarrow}^{\dagger}c_{j\downarrow}c_{j^{\prime}\uparrow},
\end{align}
where $j$ runs from 1 to the number of localized orbitals $M$ per site.
For a $t_{2g}$ shell with spin rotational invariance, we have $U_{c}=U$,
$U^{\prime}=U-2J$, $U^{\prime\prime}=U-3J$ and $J_{S}=J_{P}=J$.
In the case of the bilayer Hubbard model in the bonding-antibonding basis,
the parameters are $U_{c}=U^{\prime}=J_{S}=J_{P}=U/2$ and $U^{\prime\prime}=0$.

The Hamiltonian of the SC bath reads 
\begin{equation}
H_{\mathrm{bath}}^{\mathrm{SC}}=\sum_{k\alpha\sigma}(\epsilon_{k\alpha}-\mu)f_{k\alpha\sigma}^{\dagger}f_{k\alpha\sigma}+\sum_{k\alpha}\varDelta_{k\alpha}f_{k\alpha\uparrow}^{\dagger}f_{-k\alpha\downarrow}^{\dagger}+\sum_{k\alpha}\varDelta_{k\alpha}^{*}f_{-k\alpha\downarrow}f_{k\alpha\uparrow},
\end{equation}
where $\epsilon_{k\alpha}$ is the energy spectrum of the conduction electrons
with momentum $k$, band index $\alpha$ and spin index $\sigma=\uparrow,\downarrow$.
$\varDelta_{k\alpha}$ is the pairing amplitude. The Nambu-spinors
for the conduction electrons are defined as 
\begin{equation}
\hat{\Psi}_{k}^{\dagger}=\left[\begin{array}{ccccc}
f_{k\alpha\uparrow}^{\dagger}, & f_{-k\alpha\downarrow}, & f_{k\beta\uparrow}^{\dagger}, & f_{-k\beta\downarrow}, & \cdots\end{array}\right]\equiv\left[\begin{array}{ccc}
\hat{\Psi}_{k\alpha}^{\dagger}, & \hat{\Psi}_{k\beta}^{\dagger}, & \cdots\end{array}\right],
\end{equation}
 and $\hat{\Psi}_{k}=\left[\begin{array}{ccccc}
f_{k\alpha\uparrow}, & f_{-k\alpha\downarrow}^{\dagger}, & f_{k\beta\uparrow}, & f_{-k\beta\downarrow}^{\dagger}, & \cdots\end{array}\right]^{T}$. $H_{\mathrm{bath}}^{\mathrm{SC}}$ can be expressed in the compact form 
\begin{align}
H_{\mathrm{bath}}^{\mathrm{SC}} & =\sum_{k\alpha}\hat{\Psi}_{k\alpha}^{\dagger}\hat{E}_{k\alpha}\hat{\Psi}_{k\alpha}=\sum_{k\alpha}\left[\begin{array}{cc}
f_{k\alpha\uparrow}^{\dagger}, & f_{-k\alpha\downarrow}\end{array}\right]\left[\begin{array}{cc}
\epsilon_{k\alpha}-\mu & \varDelta_{k\alpha}\nonumber\\
\varDelta_{k\alpha}^{*} & -\epsilon_{-k\alpha}+\mu
\end{array}\right]\left[\begin{array}{c}
f_{k\alpha\uparrow}\nonumber\\
f_{-k\alpha\downarrow}^{\dagger}
\end{array}\right]\nonumber\\
 & =\sum_{k\alpha}\left[(\epsilon_{k\alpha}-\mu)f_{k\alpha\uparrow}^{\dagger}f_{k\alpha\uparrow}+(-\epsilon_{-k\alpha}+\mu)f_{-k\alpha\downarrow}f_{-k\alpha\downarrow}^{\dagger}+\varDelta_{k\alpha}f_{k\alpha\uparrow}^{\dagger}f_{-k\alpha\downarrow}^{\dagger}+\varDelta_{k\alpha}^{*}f_{-k\alpha\downarrow}f_{k\alpha\uparrow}\right]
\end{align}
with $\hat{E}_{k\alpha}=\left[\begin{array}{cc}
\epsilon_{k\alpha}-\mu & \varDelta_{k\alpha}\\
\varDelta_{k\alpha}^{*} & -\epsilon_{-k\alpha}+\mu
\end{array}\right]$.

The hybridization term in the Nambu formalism becomes 
\begin{align}
H_{\mathrm{hyb}} & =\sum_{k\alpha\sigma,j}\left[V_{k\alpha}^{j}f_{k\alpha,\sigma}^{\dagger}c_{j\sigma}+V_{k\alpha}^{j*}c_{j\sigma}^{\dagger}f_{k\alpha,\sigma}\right]\nonumber\\
 & =\sum_{k\alpha,j}\left[V_{k\alpha}^{j}f_{k\alpha,\uparrow}^{\dagger}c_{j\uparrow}+V_{k\alpha}^{j}f_{k\alpha,\downarrow}^{\dagger}c_{j\downarrow}+V_{k\alpha}^{j*}c_{j\uparrow}^{\dagger}f_{k\alpha,\uparrow}+V_{k\alpha}^{j*}c_{j\downarrow}^{\dagger}f_{k\alpha,\downarrow}\right]\nonumber\\
 & =\sum_{k\alpha,j}\left[V_{k\alpha}^{j}f_{k\alpha,\uparrow}^{\dagger}c_{j\uparrow}-V_{-k\alpha}^{j}c_{j\downarrow}f_{-k\alpha,\downarrow}^{\dagger}+V_{k\alpha}^{j*}c_{j\uparrow}^{\dagger}f_{k\alpha,\uparrow}-V_{-k\alpha}^{j*}f_{-k\alpha,\downarrow}c_{j\downarrow}^{\dagger}\right]\nonumber\\
 & =\sum_{k\alpha,j}\left[V_{k\alpha}^{j}f_{k\alpha,\uparrow}^{\dagger}c_{j\uparrow}+V_{k\alpha}^{j*}c_{j\uparrow}^{\dagger}f_{k\alpha,\uparrow}-V_{-k\alpha}^{j}c_{j\downarrow}f_{-k\alpha,\downarrow}^{\dagger}-V_{-k\alpha}^{j*}f_{-k\alpha,\downarrow}c_{j\downarrow}^{\dagger}\right]\nonumber\\
 & \equiv\hat{V}_{\uparrow}+\hat{V}_{\uparrow}^{\dagger}+\hat{V}_{\downarrow}+\hat{V}_{\downarrow}^{\dagger}.
\label{eq:Hyb_V}
\end{align}

\subsubsection*{Partition function}

In CT-HYB, we treat $H_{\mathrm{hyb}}$ as the perturbation and expand
$Z=\mathrm{Tr}e^{-\beta H}$ in terms of $H_{\mathrm{hyb}}$ as 
\begin{align}
Z & =\mathrm{Tr}\left[e^{-\beta(H_{\mathrm{loc}}+H_{\mathrm{bath}}^{\mathrm{SC}})}\mathcal{T}_{\tau}e^{-\int_{0}^{\beta}H_{\mathrm{hyb}}(\tau)d\tau}\right]\nonumber\\
 & =\sum_{n=0}^{\infty}(-1)^{n}\frac{1}{n!}\int_{0}^{\beta}d\tau_{1}\cdots\int_{0}^{\beta}d\tau_{n}\mathrm{Tr}\left[\mathcal{T}_{\tau}e^{-\beta(H_{\mathrm{loc}}+H_{\mathrm{bath}}^{\mathrm{SC}})}H_{\mathrm{hyb}}\left(\tau_{1}\right)\cdots H_{\mathrm{hyb}}\left(\tau_{n}\right)\right].
\end{align}
The particle number as well as spin conservation on the local atom
requires that the terms with non-zero contribution to $Z$ must contain
an equal number of $\hat{V}_{\sigma}$ and $\hat{V}_{\sigma}^{\dagger}$ as
\begin{equation}
Z=\sum_{n=0}^{\infty}\int_{0}^{\beta}d\tau_{1}^{\uparrow}\cdots\int_{\tau_{n-1}^{\uparrow}}^{\beta}d\tau_{n}^{\uparrow}\int_{0}^{\beta}d\tau_{1}^{\uparrow\prime}\cdots\int_{\tau_{n-1}^{\uparrow\prime}}^{\beta}d\tau_{n}^{\uparrow\prime}\sum_{m=0}^{\infty}\int_{0}^{\beta}d\tau_{1}^{\downarrow}\cdots\int_{\tau_{m-1}^{\downarrow}}^{\beta}d\tau_{m}^{\downarrow}\int_{0}^{\beta}d\tau_{1}^{\downarrow\prime}\cdots\int_{\tau_{m-1}^{\downarrow\prime}}^{\beta}d\tau_{m}^{\downarrow\prime}w_{\mathrm{trace}}
\end{equation}
 with 
\begin{equation}
w_{\mathrm{trace}}=\mathrm{Tr}\left[\mathcal{T}_{\tau}e^{-\beta(H_{\mathrm{loc}}+H_{\mathrm{bath}}^{\mathrm{SC}})}\hat{V}_{\uparrow}(\tau_{n}^{\uparrow})\hat{V}_{\uparrow}^{\dagger}(\tau_{n}^{\uparrow\prime})\cdots\hat{V}_{\uparrow}(\tau_{1}^{\uparrow})\hat{V}_{\uparrow}^{\dagger}(\tau_{1}^{\uparrow\prime})\cdot\hat{V}_{\downarrow}(\tau_{m}^{\downarrow})\hat{V}_{\downarrow}^{\dagger}(\tau_{m}^{\downarrow\prime})\cdots\hat{V}_{\downarrow}(\tau_{1}^{\downarrow})\hat{V}_{\downarrow}^{\dagger}(\tau_{1}^{\downarrow\prime})\right].
\end{equation}
After substituting $\hat{V}_{\sigma}^{(\dagger)}$ as given in Eq.~(\ref{eq:Hyb_V}) and separating
the bath and impurity operators we obtain
\begin{align}
\omega_{\mathrm{trace}}= & \sum_{k_{n}\alpha_{n},j_{n}}\sum_{k_{n}^{\prime}\alpha_{n}^{\prime},j_{n}^{\prime}}\cdots\sum_{k_{1}\alpha_{1},j_{1}}\sum_{k_{1}^{\prime}\alpha_{1}^{\prime},j_{1}^{\prime}}\sum_{\tilde{k}_{m}\tilde{\alpha}_{m},\tilde{j}_{m}}\sum_{\tilde{k}_{m}^{\prime}\tilde{\alpha}_{m}^{\prime},\tilde{j}_{m}^{\prime}}\cdots\sum_{\tilde{k}_{1}\tilde{\alpha}_{1},\tilde{j}_{1}}\sum_{\tilde{k}_{1}^{\prime}\tilde{\alpha}_{1}^{\prime},\tilde{j}_{1}^{\prime}}\nonumber\\
\times & V_{k_{n}\alpha_{n}}^{j_{n}}V_{k_{n}^{\prime}\alpha_{n}^{\prime}}^{j_{n}^{\prime}*}\cdots V_{k_{1}\alpha_{1}}^{j_{1}}V_{k_{1}^{\prime}\alpha_{1}^{\prime}}^{j_{1}^{\prime}*}V_{-\tilde{k}_{m}\tilde{\alpha}_{m}}^{\tilde{j}_{m}}V_{-\tilde{k}_{m}^{\prime}\tilde{\alpha}_{m}^{\prime}}^{\tilde{j}_{m}^{\prime}*}\cdots V_{-\tilde{k}_{1}\tilde{\alpha}_{1}}^{\tilde{j}_{1}}V_{-\tilde{k}_{1}^{\prime}\tilde{\alpha}_{1}^{\prime}}^{\tilde{j}_{1}^{\prime}*}\nonumber\\
\times & \mathrm{Tr}_{f}[\mathcal{T}_{\tau}e^{-\beta H_{\mathrm{bath}}^{\mathrm{SC}}}f_{k_{n}\alpha_{n},\uparrow}^{\dagger}(\tau_{n}^{\uparrow})f_{k_{n}^{\prime}\alpha_{n}^{\prime},\uparrow}(\tau_{n}^{\uparrow\prime})\cdots f_{k_{1}\alpha_{1},\uparrow}^{\dagger}(\tau_{1}^{\uparrow})f_{k_{1}^{\prime}\alpha_{1}^{\prime},\uparrow}(\tau_{1}^{\uparrow\prime})\nonumber\\
 & \cdot f_{-\tilde{k}_{m}\tilde{\alpha}_{m},\downarrow}^{\dagger}(\tau_{m}^{\downarrow})f_{-\tilde{k}_{m}^{\prime}\tilde{\alpha}_{m}^{\prime},\downarrow}(\tau_{m}^{\downarrow\prime})\cdots f_{-\tilde{k}_{1}\tilde{\alpha}_{1},\downarrow}^{\dagger}(\tau_{1}^{\downarrow})f_{-\tilde{k}_{1}^{\prime}\tilde{\alpha}_{1}^{\prime},\downarrow}(\tau_{1}^{\downarrow\prime})]\nonumber\\
\times & \mathrm{Tr}_{c}[\mathcal{T}_{\tau}e^{-\beta H_{\mathrm{loc}}}c_{j_{n}\uparrow}(\tau_{n}^{\uparrow})c_{j_{n}^{\prime}\uparrow}^{\dagger}(\tau_{n}^{\uparrow\prime})\cdots c_{j_{1}\uparrow}(\tau_{1}^{\uparrow})c_{j_{1}^{\prime}\uparrow}^{\dagger}(\tau_{1}^{\uparrow\prime})\nonumber\\
 & \cdot c_{\tilde{j}_{m}\downarrow}(\tau_{m}^{\downarrow})c_{\tilde{j}_{m}^{\prime}\downarrow}^{\dagger}(\tau_{m}^{\downarrow\prime})\cdots c_{\tilde{j}_{1}\downarrow}(\tau_{1}^{\downarrow})c_{\tilde{j}_{1}^{\prime}\downarrow}^{\dagger}(\tau_{1}^{\downarrow\prime})]\nonumber\\
\equiv & \frac{1}{Z_{\mathrm{bath}}^{\mathrm{SC}}}w_{\mathrm{det}}\cdot w_{\mathrm{loc}}
\end{align}
with the local trace part, 
\begin{align}
w_{\mathrm{loc}} = & \,\,\mathrm{Tr}_{c}\Big[\mathcal{T}_{\tau}e^{-\beta H_{\mathrm{loc}}}c_{j_{n}\uparrow}(\tau_{n}^{\uparrow})c_{j_{n}^{\prime}\uparrow}^{\dagger}(\tau_{n}^{\uparrow\prime})\cdots c_{j_{1}\uparrow}(\tau_{1}^{\uparrow})c_{j_{1}^{\prime}\uparrow}^{\dagger}(\tau_{1}^{\uparrow\prime})\nonumber\\
 & \times c_{\tilde{j}_{m}\downarrow}(\tau_{m}^{\downarrow})c_{\tilde{j}_{m}^{\prime}\downarrow}^{\dagger}(\tau_{m}^{\downarrow\prime})\cdots c_{\tilde{j}_{1}\downarrow}(\tau_{1}^{\downarrow})c_{\tilde{j}_{1}^{\prime}\downarrow}^{\dagger}(\tau_{1}^{\downarrow\prime})\Big],
\end{align}
and the determinant part obtained by applying Wick's theorem, 
\begin{align}
w_{\mathrm{det}}=\det\Delta\equiv\frac{1}{Z_{\mathrm{bath}}^{\mathrm{SC}}} & \sum_{k_{n}\alpha_{n}}\sum_{k_{n}^{\prime}\alpha_{n}^{\prime}}\cdots\sum_{k_{1}\alpha_{1}}\sum_{k_{1}^{\prime}\alpha_{1}^{\prime}}\sum_{\tilde{k}_{m}\tilde{\alpha}_{m}}\sum_{\tilde{k}_{m}^{\prime}\tilde{\alpha}_{m}^{\prime}}\cdots\sum_{\tilde{k}_{1}\tilde{\alpha}_{1}}\sum_{\tilde{k}_{1}^{\prime}\tilde{\alpha}_{1}^{\prime}}\\
 & V_{k_{n}\alpha_{n}}^{j_{n}}V_{k_{n}^{\prime}\alpha_{n}^{\prime}}^{j_{n}^{\prime}*}\cdots V_{k_{1}\alpha_{1}}^{j_{1}}V_{k_{1}^{\prime}\alpha_{1}^{\prime}}^{j_{1}^{\prime}*}V_{-\tilde{k}_{m}\tilde{\alpha}_{m}}^{\tilde{j}_{m}}V_{-\tilde{k}_{m}^{\prime}\tilde{\alpha}_{m}^{\prime}}^{\tilde{j}_{m}^{\prime}*}\cdots V_{-\tilde{k}_{1}\tilde{\alpha}_{1}}^{\tilde{j}_{1}}V_{-\tilde{k}_{1}^{\prime}\tilde{\alpha}_{1}^{\prime}}^{\tilde{j}_{1}^{\prime}*}\nonumber\\
 & \mathrm{Tr}_{f}\Big[\mathcal{T}_{\tau}e^{-\beta H_{\mathrm{bath}}^{\mathrm{SC}}}f_{k_{n}\alpha_{n},\uparrow}^{\dagger}(\tau_{n}^{\uparrow})f_{k_{n}^{\prime}\alpha_{n}^{\prime},\uparrow}(\tau_{n}^{\uparrow\prime})\cdots f_{k_{1}\alpha_{1},\uparrow}^{\dagger}(\tau_{1}^{\uparrow})f_{k_{1}^{\prime}\alpha_{1}^{\prime},\uparrow}(\tau_{1}^{\uparrow\prime})\nonumber\\
 & \times f_{-\tilde{k}_{m}\tilde{\alpha}_{m},\downarrow}^{\dagger}(\tau_{m}^{\downarrow})f_{-\tilde{k}_{m}^{\prime}\tilde{\alpha}_{m}^{\prime},\downarrow}(\tau_{m}^{\downarrow\prime})\cdots f_{-\tilde{k}_{1}\tilde{\alpha}_{1},\downarrow}^{\dagger}(\tau_{1}^{\downarrow})f_{-\tilde{k}_{1}^{\prime}\tilde{\alpha}_{1}^{\prime},\downarrow}(\tau_{1}^{\downarrow\prime})\Big],
\end{align}
 where $Z_{\mathrm{bath}}^{\mathrm{SC}}=\mathrm{Tr}_{f}\big[\mathcal{T}_{\tau}e^{-\beta H_{\mathrm{bath}}^{\mathrm{SC}}}\big]$.
In general, $\Delta$ is a non-block diagonal matrix with non-zero elements
between two different orbital indices. In an appropriate basis, $\Delta$
can become block diagonal. In the following, we consider the situation
where $\Delta$ is block diagonal in the orbital index, which is the
case for the bilayer Hubbard model, 
\begin{equation}
\det\Delta=\prod_{j=1}^{N}\det\Delta_{j}.
\end{equation}
A certain configuration in $Z$ contains $n_{j\sigma}^{\prime}$ local
creation operators $\{c_{j\sigma}^{\dagger}(\tau_{1}^{(j,\sigma)\prime}),\cdots,c_{j\sigma}^{\dagger}(\tau_{n_{j\sigma}^{\prime}}^{(j,\sigma)\prime})\}$
and $n_{j\sigma}$ local annihilation operators $\{c_{j\sigma}(\tau_{1}^{(j,\sigma)}),\cdots,c(\tau_{n_{j\sigma}}^{(j,\sigma)})\}$
for orbital $j$ and spin $\sigma$. There are four types of hybridization
matrix elements in $\Delta_{j}$. The normal elements for $j$ and $\uparrow$
read
\begin{equation}
\Delta_{\mathrm{nor}}^{(j\uparrow\uparrow)}(\tau_{l}^{(j\uparrow)\prime}-\tau_{m}^{(j\uparrow)})=\sum_{k_{l}^{\prime}\alpha_{l}^{\prime}}\sum_{k_{m}\alpha_{m}}V_{k_{l}^{\prime}\alpha_{l}^{\prime}}^{j*}V_{k_{m}\alpha_{m}}^{j}\mathrm{Tr}_{f}\Big[\mathcal{T}_{\tau}e^{-\beta H_{\mathrm{bath}}^{\mathrm{SC}}}f_{k_{m}\alpha_{m},\uparrow}^{\dagger}(\tau_{m}^{(j\uparrow)})f_{k_{l}^{\prime}\alpha_{l}^{\prime},\uparrow}(\tau_{l}^{(j\uparrow)\prime})\Big],
\end{equation}
while for $j$ and $\downarrow$  they are
\begin{equation}
-[\Delta_{\mathrm{nor}}^{(j\downarrow\downarrow)}[-(\tau_{l}^{(j\downarrow)\prime}-\tau_{m}^{(j\downarrow)})]=\sum_{k_{l}^{\prime}\alpha_{l}^{\prime}}\sum_{k_{m}\alpha_{m}}V_{-k_{l}^{\prime}\alpha_{l}^{\prime}}^{j*}V_{-k_{m}\alpha_{m}}^{j}\mathrm{Tr}_{f}\Big[\mathcal{T}_{\tau}e^{-\beta H_{\mathrm{bath}}^{\mathrm{SC}}}f_{-k_{m}\alpha_{m},\downarrow}^{\dagger}(\tau_{m}^{(j\downarrow)})f_{-k_{l}^{\prime}\alpha_{l}^{\prime},\downarrow}(\tau_{l}^{(j\downarrow)\prime})\Big].
\end{equation}
The anomalous element for $j\uparrow j\downarrow$ reads
\begin{equation}
\Delta_{\mathrm{ano}}^{(j\uparrow\downarrow)}(\tau_{l}^{(j\uparrow)\prime}-\tau_{m}^{(j\downarrow)\prime})=\sum_{k_{l}^{\prime}\alpha_{l}^{\prime}}\sum_{k_{m}^{\prime}\alpha_{m}^{\prime}}V_{k_{l}^{\prime}\alpha_{l}^{\prime}}^{j*}V_{-k_{m}^{\prime}\alpha_{m}^{\prime}}^{j*}\mathrm{Tr}_{f}\Big[\mathcal{T}_{\tau}e^{-\beta H_{\mathrm{bath}}^{\mathrm{SC}}}f_{-k_{m}^{\prime}\alpha_{m}^{\prime},\downarrow}(\tau_{m}^{(j\downarrow)\prime})f_{k_{l}^{\prime}\alpha_{l}^{\prime},\uparrow}(\tau_{l}^{(j\uparrow)\prime})\Big],
\end{equation}
 and its counterpart for $j\downarrow j\uparrow$ is
\begin{equation}
\Delta_{\mathrm{ano}}^{(j\downarrow\uparrow)}(\tau_{l}^{(j\downarrow)}-\tau_{m}^{(j\uparrow)})=\sum_{k_{l}\alpha_{l}}\sum_{k_{m}\alpha_{m}}V_{-k_{l}\alpha_{l}}^{j}V_{k_{m}\alpha_{m}}^{j}\mathrm{Tr}_{f}\Big[\mathcal{T}_{\tau}e^{-\beta H_{\mathrm{bath}}^{\mathrm{SC}}}f_{k_{m}\alpha_{m},\uparrow}^{\dagger}(\tau_{m}^{(j\uparrow)})f_{-k_{l}\alpha_{l},\downarrow}^{\dagger}(\tau_{l}^{(j\downarrow)})\Big].
\end{equation}
$\Delta_{j}$, a $(n_{j\uparrow}^{\prime}+n_{j\downarrow})\times(n_{j\uparrow}+n_{j\downarrow}^{\prime})$ matrix,
can be expressed as 
\begin{equation}
\Delta_{j}=\left[\begin{array}{cccccc}
\Delta_{\mathrm{nor}}^{(j\uparrow\uparrow)}(\tau_{1}^{(j\uparrow)\prime}-\tau_{1}^{(j\uparrow)}) & \cdots & \Delta_{\mathrm{nor}}^{(j\uparrow\uparrow)}(\tau_{1}^{(j\uparrow)\prime}-\tau_{n_{j\uparrow}}^{(j\uparrow)}) & \Delta_{\mathrm{ano}}^{(j\uparrow\downarrow)}(\tau_{1}^{(j\uparrow)\prime}-\tau_{1}^{(j\downarrow)\prime}) & \cdots & \Delta_{\mathrm{ano}}^{(j\uparrow\downarrow)}(\tau_{1}^{(j\uparrow)\prime}-\tau_{n_{j\downarrow}^{\prime}}^{(j\downarrow)\prime})\\
\vdots & \vdots & \vdots & \vdots & \vdots & \vdots\\
\Delta_{\mathrm{nor}}^{(j\uparrow\uparrow)}(\tau_{n_{j\uparrow}^{\prime}}^{(j\uparrow)\prime}-\tau_{1}^{(j\uparrow)}) & \cdots & \Delta_{\mathrm{nor}}^{(j\uparrow\uparrow)}(\tau_{n_{j\uparrow}^{\prime}}^{(j\uparrow)\prime}-\tau_{n_{j\uparrow}}^{(j\uparrow)}) & \Delta_{\mathrm{ano}}^{(j\uparrow\downarrow)}(\tau_{n_{j\uparrow}^{\prime}}^{(j\uparrow)\prime}-\tau_{1}^{(j\downarrow)\prime}) & \cdots & \Delta_{\mathrm{ano}}^{(j\uparrow\downarrow)}(\tau_{n_{j\uparrow}^{\prime}}^{(j\uparrow)\prime}-\tau_{n_{j\downarrow}^{\prime}}^{(j\downarrow)\prime})\\
\Delta_{\mathrm{ano}}^{(j\downarrow\uparrow)}(\tau_{1}^{(j\downarrow)}-\tau_{1}^{(j\uparrow)}) & \cdots & \Delta_{\mathrm{ano}}^{(j\downarrow\uparrow)}(\tau_{1}^{(j\downarrow)}-\tau_{n_{j\uparrow}}^{(j\uparrow)}) & -\Delta_{\mathrm{nor}}^{(j\downarrow\downarrow)}[-(\tau_{1}^{(j\downarrow)}-\tau_{1}^{(j\downarrow)})] & \cdots & -\Delta_{\mathrm{nor}}^{(j\downarrow\downarrow)}[-(\tau_{1}^{(j\downarrow)\prime}-\tau_{n_{j\downarrow}^{\prime}}^{(j\downarrow)})]\\
\vdots & \vdots & \vdots & \vdots & \vdots & \vdots\\
\Delta_{\mathrm{ano}}^{(j\downarrow\uparrow)}(\tau_{n_{j\downarrow}}^{(j\downarrow)}-\tau_{1}^{(j\uparrow)}) & \cdots & \Delta_{\mathrm{ano}}^{(j\downarrow\uparrow)}(\tau_{n_{j\downarrow}}^{(j\downarrow)}-\tau_{n_{j\uparrow}}^{(j\uparrow)}) & -\Delta_{\mathrm{nor}}^{(j\downarrow\downarrow)}[-(\tau_{n_{j\downarrow}^{\prime}}^{(j\downarrow)\prime}-\tau_{1}^{(j\downarrow)})] & \cdots & -\Delta_{\mathrm{nor}}^{(j\downarrow\downarrow)}[-(\tau_{n_{j\downarrow}^{\prime}}^{(j\downarrow)\prime}-\tau_{n_{j\downarrow}^{\prime}}^{(j\downarrow)})]
\end{array}\right].
\end{equation}
$\Delta_{j}$ is a square matrix with $n_{j\uparrow}^{\prime}+n_{j\downarrow}=n_{j\uparrow}+n_{j\downarrow}^{\prime}$
but {\it not} necessarily $n_{j\sigma}=n_{j\sigma}^{\prime}$. This means
the number of creation operators for spin-orbital index $j\sigma$
can be different from the number of destruction operators
for the same index. In other words, if we write the hybridization matrix in a block form representing the normal and anomalous components, $\Delta_{j}=\left[\begin{array}{cc}
A & B\\
C & D
\end{array}\right]$, each submatrix can be 
a non-square matrix. 
In the end,
the partition function reads

\begin{align}
Z & =Z_{\mathrm{bath}}^{\mathrm{SC}}\left[\prod_{j=1}^{M}\prod_{\sigma=\uparrow,\downarrow}\sum_{n_{j\sigma},n_{j\sigma}^{\prime}=0}^{\infty}\int_{0}^{\beta}d\tau_{1}^{(j\sigma)}\cdots\int_{\tau_{n-1}^{(j\sigma)}}^{\beta}d\tau_{n^{(j\sigma)}}^{(j\sigma)}\int_{0}^{\beta}d\tau_{1}^{(j\sigma)\prime}\cdots\int_{\tau_{n-1}^{(j\sigma)\prime}}^{\beta}d\tau_{n^{(j\sigma)\prime}}^{(j\sigma)\prime}\det\Delta_{j}\right]\nonumber\\
 & \times\mathrm{Tr}_{c}\Big[\mathcal{T}_{\tau}e^{-\beta H_{\mathrm{loc}}}\prod_{j=1}^{M}c_{j\uparrow}(\tau_{n_{j\uparrow}}^{(j\uparrow)})c_{j\uparrow}^{\dagger}(\tau_{n_{j\uparrow}^{\prime}}^{(j\uparrow)\prime})\cdots c_{j\uparrow}(\tau_{1}^{(j\uparrow)})c_{j\uparrow}^{\dagger}(\tau_{1}^{(j\uparrow)\prime})\times c_{j\downarrow}(\tau_{n_{j\downarrow}}^{(j\downarrow)})c_{j\downarrow}^{\dagger}(\tau_{n_{j\downarrow}^{\prime}}^{(j\downarrow)\prime})\cdots c_{j\downarrow}(\tau_{1}^{(j\downarrow)})c_{j\downarrow}^{\dagger}(\tau_{1}^{(j\downarrow)\prime})\Big]s_{c},
\end{align}
where $s_{c}$ is the permutation sign from grouping $\{c,c^{\dagger}\}$
 operators by their orbital indices for sake of clearness.

\subsubsection*{Updates within $\mathcal{C}_Z$}

The normal pair insertion/removal update is a simple and necessary
update, which involves one creation operator $c_{j\sigma}^{\dagger}(\tau^{(j\sigma)})$
and one annihilation operator $c_{j\sigma}(\tau^{(j\sigma)})$ for the
same spin-orbital $j\sigma$. This update changes the expansion order
by $\pm 1$ as $n_{j\sigma}^{(\prime)}\rightarrow n_{j\sigma}^{(\prime)}\pm1$.
If $\sigma=\uparrow$ ($=\downarrow)$ the update will modify the block matrices $A$,
$C$ and $B$ ($D$, $B$, $C$ ) in $\Delta_{j}$.

As pointed out by S$\acute{\text{e}}$mon {et al.} \cite{Semon2014}, four-operator
(termed $4$-op) updates are also necessary for ergodicity in the case of superconducting
states. This can be seen by considering the simple configuration, 
\begin{equation}
\beta|-c_{j^{\prime}\uparrow}^{\dagger}(\tau_{1}^{(j^{\prime}\uparrow)\prime})-c_{j^{\prime}\downarrow}^{\dagger}(\tau_{1}^{(j^{\prime}\downarrow)\prime})-c_{j\downarrow}(\tau_{1}^{(j\downarrow)})-c_{j\uparrow}(\tau_{1}^{(j\uparrow)})-|0,
\end{equation}
where the order of $\{\tau\}$ can be arbitrary and $j\ne j^\prime$. This configuration
cannot be generated by two successive insertion updates, since the
local trace after the first insertion is zero. The corresponding determinant
is 
\begin{equation}
w_{\mathrm{det}}=\Delta_{\mathrm{ano}}^{(j\downarrow\uparrow)}(\tau_{1}^{(j\downarrow)}-\tau_{1}^{(j\uparrow)})\times\Delta_{\mathrm{ano}}^{(j^{\prime}\uparrow\downarrow)}(\tau_{1}^{(j^{\prime}\uparrow)\prime}-\tau_{1}^{(j^{\prime}\downarrow)\prime}),
\end{equation}
and the local trace  
\begin{equation}
w_{\mathrm{loc}}=\mathrm{Tr}_{c}[\mathcal{T}_{\tau}e^{-\beta H_{\mathrm{loc}}}c_{j^{\prime}\uparrow}^{\dagger}(\tau_{1}^{(j^{\prime}\uparrow)\prime})c_{j^{\prime}\downarrow}^{\dagger}(\tau_{1}^{(j^{\prime}\downarrow)\prime})c_{j\downarrow}(\tau_{1}^{(j\downarrow)})c_{j\uparrow}(\tau_{1}^{(j\uparrow)})]\ne0 
\end{equation}
 if $J_{P}\ne0$.

Starting from an arbitrary configuration, the 4-op insertion/removal update 
changes the expansion order as $n_{j\sigma}\rightarrow n_{j\sigma}\pm1$
and $n_{j^{\prime}\sigma}^{\prime}\rightarrow n_{j^{\prime}\sigma}^{\prime}\pm1$.
It changes the matrices $C$, $A$, $D$ in $\Delta_{j}$ and the matrices $B$,
$D$, $A$ in $\Delta_{j^{\prime}}$.

\subsection*{Green's function}

The normal ($G$) and anomalous ($F$) single-particle Green's functions are obtained
by inserting two corresponding operators in the local trace part of
$Z$ as
\begin{align}
G_{j\uparrow\uparrow}(\tau^{(j\uparrow)},\tau^{(j\uparrow)\prime})= & -\langle\mathcal{T}_{\tau}c_{j\uparrow}(\tau^{(j\uparrow)})c_{j\uparrow}^{\dagger}(\tau^{(j\uparrow)\prime})\rangle\nonumber\\
= & Z_{\mathrm{bath}}^{\mathrm{SC}}\left[\prod_{j=1}^{M}\prod_{\sigma=\uparrow,\downarrow}\sum_{n_{j\sigma},n_{j\sigma}^{\prime}=0}^{\infty}\int_{0}^{\beta}d\tau_{1}^{(j\sigma)}\cdots\int_{\tau_{n-1}^{(j\sigma)}}^{\beta}d\tau_{n^{(j\sigma)}}^{(j\sigma)}\int_{0}^{\beta}d\tau_{1}^{(j\sigma)\prime}\cdots\int_{\tau_{n-1}^{(j\sigma)\prime}}^{\beta}d\tau_{n^{(j\sigma)\prime}}^{(j\sigma)\prime}\det\Delta_{j}\right]\nonumber\\
\times & \mathrm{Tr}_{c}[\mathcal{T}_{\tau}e^{-\beta H_{\mathrm{loc}}}\prod_{j=1}^{M}c_{j\uparrow}(\tau_{n_{j\uparrow}}^{(j\uparrow)})c_{j\uparrow}^{\dagger}(\tau_{n_{j\uparrow}^{\prime}}^{(j\uparrow)\prime})\cdots c_{j\uparrow}(\tau_{1}^{(j\uparrow)})c_{j\uparrow}^{\dagger}(\tau_{1}^{(j\uparrow)\prime})\nonumber\\
\times & c_{j\downarrow}(\tau_{n_{j\downarrow}}^{(j\downarrow)})c_{j\downarrow}^{\dagger}(\tau_{n_{j\downarrow}^{\prime}}^{(j\downarrow)\prime})\cdots c_{j\downarrow}(\tau_{1}^{(j\downarrow)})c_{j\downarrow}^{\dagger}(\tau_{1}^{(j\downarrow)\prime})c_{j\uparrow}(\tau^{(j\uparrow)})c_{j\uparrow}^{\dagger}(\tau^{(j\uparrow)\prime})]s_{c},
\end{align}
and 
\begin{align}
F_{j\uparrow\downarrow}(\tau^{(j\uparrow)},\tau^{(j\downarrow)})= & -\langle\mathcal{T}_{\tau}c_{j\uparrow}(\tau^{(j\uparrow)})c_{j\downarrow}(\tau^{(j\downarrow)})\rangle\nonumber\\
= & Z_{\mathrm{bath}}^{\mathrm{SC}}\left[\prod_{j=1}^{M}\prod_{\sigma=\uparrow,\downarrow}\sum_{n_{j\sigma},n_{j\sigma}^{\prime}=0}^{\infty}\int_{0}^{\beta}d\tau_{1}^{(j\sigma)}\cdots\int_{\tau_{n-1}^{(j\sigma)}}^{\beta}d\tau_{n^{(j\sigma)}}^{(j\sigma)}\int_{0}^{\beta}d\tau_{1}^{(j\sigma)\prime}\cdots\int_{\tau_{n-1}^{(j\sigma)\prime}}^{\beta}d\tau_{n^{(j\sigma)\prime}}^{(j\sigma)\prime}\det\Delta_{j}\right]\nonumber\\
\times & \mathrm{Tr}_{c}[\mathcal{T}_{\tau}e^{-\beta H_{\mathrm{loc}}}\prod_{j=1}^{M}c_{j\uparrow}(\tau_{n_{j\uparrow}}^{(j\uparrow)})c_{j\uparrow}^{\dagger}(\tau_{n_{j\uparrow}^{\prime}}^{(j\uparrow)\prime})\cdots c_{j\uparrow}(\tau_{1}^{(j\uparrow)})c_{j\uparrow}^{\dagger}(\tau_{1}^{(j\uparrow)\prime})\nonumber\\
\times & c_{j\downarrow}(\tau_{n_{j\downarrow}}^{(j\downarrow)})c_{j\downarrow}^{\dagger}(\tau_{n_{j\downarrow}^{\prime}}^{(j\downarrow)\prime})\cdots c_{j\downarrow}(\tau_{1}^{(j\downarrow)})c_{j\downarrow}^{\dagger}(\tau_{1}^{(j\downarrow)\prime})c_{j\uparrow}(\tau^{(j\uparrow)})c_{j\downarrow}(\tau^{(j\downarrow)})]s_{c}.
\end{align}
We only consider the paramagnetic phase and therefore $G_{j\downarrow\downarrow}(\tau)$
and $F_{j\downarrow\uparrow}(\tau)$ can be obtained from $G_{j\uparrow\uparrow}(\tau)$
and $F_{j\uparrow\downarrow}(\tau)$ using time-reversal
symmetry: $G_{j\downarrow\downarrow}(\tau)=G_{j\uparrow\uparrow}(\tau)$,
$F_{j\downarrow\uparrow}(\tau)=F_{j\uparrow\downarrow}(\tau)^{*}$.
Since the partition function in the Nambu formalism has a similar structure as the partition function in the conventional CT-HYB
for the normal phase, we can directly write down the conventional
estimator to measure $G_{j\uparrow\uparrow}$ and $F_{j\uparrow\downarrow}$ as
\begin{equation}
\begin{array}{l}
G_{j\uparrow\uparrow}\left(\tau-\tau^{\prime}\right)=-\frac{1}{\beta}\left\langle \sum_{lm=1}^{(n_{j\uparrow}^{\prime}+n_{j\downarrow})}(\Delta_{j}^{-1})_{lm}\delta^{-}\left(\tau-\tau^{\prime},\tau_{m}-\tau_{l}\right)\delta_{j\uparrow,m}\delta_{j\uparrow,l}\right\rangle _{\mathrm{MC}}\end{array},
\label{eq:G_conv}
\end{equation}
\begin{equation}
F_{j\uparrow\downarrow}\left(\tau-\tau^{\prime}\right)=-\frac{1}{\beta}\left\langle \sum_{lm=1}^{(n_{j\uparrow}^{\prime}+n_{j\downarrow})}(\Delta_{j}^{-1})_{lm}\delta^{-}\left(\tau-\tau^{\prime},\tau_{m}-\tau_{l}\right)\delta_{j\uparrow,m}\delta_{j\downarrow,l}\right\rangle _{\mathrm{MC}}.
\label{eq:F_conv}
\end{equation}
Equation (\ref{eq:G_conv}) [Eq.~(\ref{eq:F_conv})] is obtained by removing the normal (anomalous) hybridization lines
attached to two operators ($c_{j\uparrow}$ and $c_{j\uparrow}^{\dagger}$
to measure $G$, $c_{j\uparrow}$ and $c_{j\downarrow}$ to measure
$F$) in a certain configuration of $Z$.

The conventional estimator yields bad statistics if the average expansion
order becomes small $\langle n_{j\sigma}\rangle\rightarrow0$ and it cannot be applied
if $\langle n_{j\sigma}\rangle=0$ (where 
$\langle n_{j\sigma}\rangle$ should not be confused with the average occupation number $\langle\hat{n}_{j\sigma}\rangle$).
This happens when the hybridization function reaches the atomic limit
$\Delta^{(j\sigma)}\rightarrow0$. For example, it appears in the
Falicov-Kimball model \cite{FKmodel}. As shown in the main text, this situation also appears
in the bilayer Hubbard model when the narrow band reaches the flat-band
limit ($W_\beta\rightarrow0$), since the hybridization strength is proportional to its band width $V_\beta\propto W_\beta\rightarrow0$.

\subsection*{Normal worm sampling}

In the normal worm sampling, one treats $\text{\ensuremath{c_{j\uparrow}}(\ensuremath{\tau^{(j\uparrow)}})}$
and $c_{j\uparrow}^{\dagger}(\tau^{(j\uparrow)\prime})$ as the worm
operators, and includes the worm space $\mathcal{C}_{G_{j\uparrow\uparrow}^{(1)}}$.
The modified partition function is

\begin{align}
Z_{{G_{j\uparrow\uparrow}^{(1)}}}=\int d\tau^{(j\uparrow)}d\tau^{(j\uparrow)\prime}G_{j\uparrow\uparrow}(\tau^{(j\uparrow)},\tau^{(j\uparrow)\prime}).
\end{align}

\subsubsection*{Updates within $\mathcal{C}_{G_{j\uparrow\uparrow}^{(1)}}$}

The updates within $\mathcal{C}_{G_{j\uparrow\uparrow}^{(1)}}$ are
analogous to those in $\mathcal{C}_{Z}$. The only difference 
is that there are no hybridization lines attached to the
worm operators. The necessary updates are normal pair insertion/removal
and 4-op updates. Furthermore, we also implemented worm shift/replacement
updates to reduce the auto-correlation time.

\subsubsection*{Updates between $\mathcal{C}_{Z}$ and $\mathcal{C}_{G_{j\uparrow\uparrow}^{(1)}}$}

In the worm insertion update, we start from a random configuration
in $\mathcal{C}_{Z}$ 
and randomly choose
two imaginary times for the worm operators $\text{\ensuremath{c_{j\uparrow}}(\ensuremath{\tau^{(j\uparrow)}})}$
and $c_{j\uparrow}^{\dagger}(\tau^{(j\uparrow)\prime})$. The worm
removal update is the inverse process which removes the worm
operators.

The Metropolis acceptance rates for the worm insertion and removal updates
are 
\begin{equation}
p(\mathcal{C}_{Z}\rightarrow\mathcal{C}_{G_{j\uparrow\uparrow}^{(1)}})=\min\left[1,\eta_{G^{(1)}}\cdot\frac{w_{\mathrm{loc}}(\{\tau\}_{Z},\tau^{(j\uparrow)},\tau^{(j^{\prime}\uparrow)\prime})}{w_{\mathrm{loc}}(\{\tau\}_{Z})}\beta^{2}\right],
\end{equation}
and 
\begin{equation}
p(\mathcal{C}_{G_{j\uparrow\uparrow}^{(1)}}\rightarrow\mathcal{C}_{Z})=\min\left[1,\frac{1}{\eta_{G^{(1)}}}\cdot\frac{w_{\mathrm{loc}}(\{\tau\}_{Z})}{w_{\mathrm{loc}}(\{\tau\}_{Z},\tau^{(j\uparrow)},\tau^{(j^{\prime}\uparrow)\prime})}\cdot\frac{1}{\beta^{2}}\right],
\end{equation}
respectively.

\subsubsection*{Worm measurement}

The measurement formula for the anomalous Green's function is
\begin{equation}
G_{{F^{(1)}}}^{(1)}(\tau)=\frac{1}{\eta_{G^{(1)}}}\frac{N_{G^{(1)}}}{N_{Z}}\frac{\left\langle \mathrm{sgn}(\mathcal{C}_{G^{(1)}})\cdot\delta\left(\tau,\tau_{i}-\tau_{j}\right)\right\rangle _{\mathrm{MC}}}{\langle\mathrm{sgn}(\text{\ensuremath{\mathcal{C}_{Z}})}\rangle_{\mathrm{MC}}},
\label{eq:G_worm}
\end{equation}
where $N_{G^{(1)}}$ ($\text{\ensuremath{N_{Z}}}$) is the number
of Monte Carlo steps taken in $\mathcal{C}_{G^{(1)}}$ ($\mathcal{C}_{Z}$), 
$\text{\ensuremath{\mathrm{sgn}}(\ensuremath{\mathcal{C}_{G^{(1)}}})}$
is the sign of a certain configuration of $\mathcal{C}_{G^{(1)}}$,
and $\langle\mathrm{sgn}(\text{\ensuremath{\mathcal{C}_{Z}})}\rangle_{\mathrm{MC}}$
is the average sign of the configurations in the $\mathcal{C}_Z$ space.

\subsection*{Anomalous worm sampling}

In the anomalous worm sampling, one treats $\text{\ensuremath{c_{j\uparrow}}(\ensuremath{\tau^{(j\uparrow)}})}$
and $c_{j\downarrow}(\tau^{(j\downarrow)})$ as the worm operators,
and considers the worm space $\mathcal{C}_{F_{j\uparrow\downarrow}^{(1)}}$.
The modified partition function is
\begin{align}
Z_{{F_{j\uparrow\downarrow}^{(1)}}}=\int d\tau^{(j\uparrow)}d\tau^{(j\downarrow)}F_{j\uparrow\downarrow}(\tau^{(j\uparrow)},\tau^{(j\downarrow)}).
\end{align}

\subsubsection*{Updates within $\mathcal{C}_{F_{j\uparrow\downarrow}^{(1)}}$}

Due to the existence of two worm annihilation operators, a non-zero local trace
$w_{\mathrm{loc}}$ in $F$ requires additionally two creation
operators $c_{j^{\prime}\uparrow}^{\dagger}$ and $c_{j^{\prime}\downarrow}^{\dagger}$.
Thus the first non-zero diagram in $F$ contains two worm operators
and two normal creation operators. Then we can perform normal pair insertion
updates starting from this diagram, and normal pair removal updates for diagrams
with more than these four operators. The 4-op update used in $Z$ is
also necessary to reach ergodicity within the $F$ space. Such updates can
result in a non-zero local trace and non-zero determinant even though the final configuration cannot be reached by two successive normal pair insertion/removal
updates. As in the normal worm sampling, worm shift/replacement
updates allow to reduce the auto-correlation time.

\subsubsection*{Updates between $\mathcal{C}_{Z}$ and $\mathcal{C}_{F_{j\uparrow\downarrow}^{(1)}}$}

The anomalous worm insertion update is proposed as follows. We start
from a random configuration $\mathcal{C}_{Z}$ in the partition function
space and randomly choose two imaginary times for the worm operators
$\text{\ensuremath{c_{j\uparrow}}(\ensuremath{\tau^{(j\uparrow)}})}$
and $c_{j\downarrow}(\tau^{(j\downarrow)})$. Then we randomly assign
an orbital index $j^{\prime}\in[1,2,\cdots,M]$ and randomly pick the imaginary
times for the two normal operators $c_{j^{\prime}\uparrow}^{\dagger}(\tau^{(j^{\prime}\uparrow)\prime})$
and $c_{j^{\prime}\downarrow}^{\dagger}(\tau^{(j^{\prime}\downarrow)\prime})$.
We suppose the number of $c_{j^{\prime}\uparrow}^{\dagger}$ ($c_{j^{\prime}\uparrow}^{\dagger}$) operators
is $n_{j^{\prime}\uparrow}^{\prime}$ ($n_{j^{\prime}\downarrow}^{\prime}$)
in $\mathcal{C}_{Z}$ before the update. The anomalous worm removal
update is performed by randomly choosing an orbital index $j^{\prime}$
and then randomly selecting one of the existing $c_{j^{\prime}\uparrow}^{\dagger}$
and $c_{j^{\prime}\uparrow}^{\dagger}$.

The Metropolis acceptance rates for the worm insertion and removal updates
are 
\begin{equation}
p(\mathcal{C}_{Z}\rightarrow\mathcal{C}_{F_{j\uparrow\downarrow}^{(1)}})=\min\left[1,\eta_{F^{(1)}}\cdot\frac{\det\Delta_{j^{\prime}}(\{\tau^{j^{\prime}}\}_{Z},\tau^{(j^{\prime}\uparrow)\prime},\tau^{(j^{\prime}\downarrow)\prime})}{\det\Delta_{j^{\prime}}(\{\tau^{j^{\prime}}\}_{Z})}\cdot\frac{w_{\mathrm{loc}}(\{\tau\}_{Z},\tau^{(j\uparrow)},\tau^{(j\downarrow)},\tau^{(j^{\prime}\uparrow)\prime},\tau^{(j^{\prime}\downarrow)\prime})}{w_{\mathrm{loc}}(\{\tau\}_{Z})}\cdot\frac{\beta^{4}}{(n_{j^{\prime}\uparrow}^{\prime}+1)(n_{j^{\prime}\downarrow}^{\prime}+1)}\right]
\label{eq:ZtoF}
\end{equation}
and
\begin{equation}
p(\mathcal{C}_{F_{j\uparrow\downarrow}^{(1)}}\rightarrow\mathcal{C}_{Z})=\min\left[1,\frac{1}{\eta_{F^{(1)}}}\cdot\frac{\det\Delta_{j^\prime}(\{\tau^{j^\prime}\}_{Z})}{\det\Delta_{j^\prime}(\{\tau^{j^\prime}\}_{Z},\tau^{(j^{\prime}\uparrow)\prime},\tau^{(j^{\prime}\downarrow)\prime})}\cdot\frac{w_{\mathrm{loc}}(\{\tau\}_{Z})}{w_{\mathrm{loc}}(\{\tau\}_{Z},\tau^{(j\uparrow)},\tau^{(j\downarrow)},\tau^{(j^{\prime}\uparrow)\prime},\tau^{(j^{\prime}\downarrow)\prime})}\cdot\frac{n_{j^{\prime}\uparrow}^{\prime}n_{j^{\prime}\downarrow}^{\prime}}{\beta^{4}}\right],
\label{eq:FtoZ}
\end{equation}
respectively. The determinant ratios in Eqs.~(\ref{eq:ZtoF}) and (\ref{eq:FtoZ}) appear because of the two normal operators which are inserted/removed with the worm operators.

\subsubsection*{Worm Measurement}

The measurement formula for the anomalous Green's function is
\begin{equation}
G_{{F^{(1)}}}^{(1)}(\tau)=\frac{1}{\eta_{F^{(1)}}}\frac{N_{F^{(1)}}}{N_{Z}}\frac{\left\langle \mathrm{sgn}(\mathcal{C}_{F^{(1)}})\cdot\delta\left(\tau,\tau_{i}-\tau_{j}\right)\right\rangle _{\mathrm{MC}}}{\langle\mathrm{sgn}(\text{\ensuremath{\mathcal{C}_{Z}})}\rangle_{\mathrm{MC}}},
\label{eq:F_worm}
\end{equation}
where $N_{F^{(1)}}$ is the number of Monte Carlo steps taken
in $\mathcal{C}_{F^{(1)}}$, and $\text{\ensuremath{\mathrm{sgn}}(\ensuremath{\mathcal{C}_{F^{(1)}}})}$
is the sign of a certain configuration in $\mathcal{C}_{F^{(1)}}$.

\subsection*{Tests of the worm sampling code}

We first benchmark the worm sampling in a parameter region where the worm
sampling is not necessary. In the bilayer Hubbard model, the hybridization
function in the anti-bonding orbital $\beta$ is nonzero if $W_{\beta}/W_{\alpha} > 0.$ 
We compare in Fig.~\ref{fig:GF_compare} the normal and anomalous Green's
functions measured by the conventional estimator (blue lines) according to Eqs.~(\ref{eq:G_conv},\ref{eq:F_conv})
and with worm sampling (blue lines) according to Eqs.~(\ref{eq:G_worm},\ref{eq:F_worm}).
For $W_{\beta}/W_{\alpha}$=1, there is a strong hybridization in the $\beta$ orbital and the conventional
estimator is more efficient than worm sampling, see panel (a). 
As one reduces the narrow band width to $W_\beta/W_\alpha=0.2$ (panel (b)), the hybridization strength in the $\beta$ band becomes much 
smaller. As a result, the conventional measurement of $G_{\beta\uparrow\uparrow}(\tau)$ and especially $F_{\beta\uparrow\downarrow}(\tau)$ contains
a lot of noise, which can be significantly reduced by applying the normal and anomalous worm sampling. 
As we approach the narrow-band limit for $W_\beta/W_\alpha=0.05$ (panel (c)), the noise in the conventional estimate of $F_{\beta\uparrow\downarrow}(\tau)$ becomes
very severe. 
When the anti-bonding band is flat (panel (d)), the conventional estimator (which measures zero 
when the expansion order $n_{\beta\sigma}=n_{\beta\sigma}^\prime=0$)
cannot be used anymore due to a zero hybridization, while the worm sampling still allows to measure the normal and anomalous Green's functions.

\begin{figure*}
\includegraphics[clip,width=5.5in,angle=0]{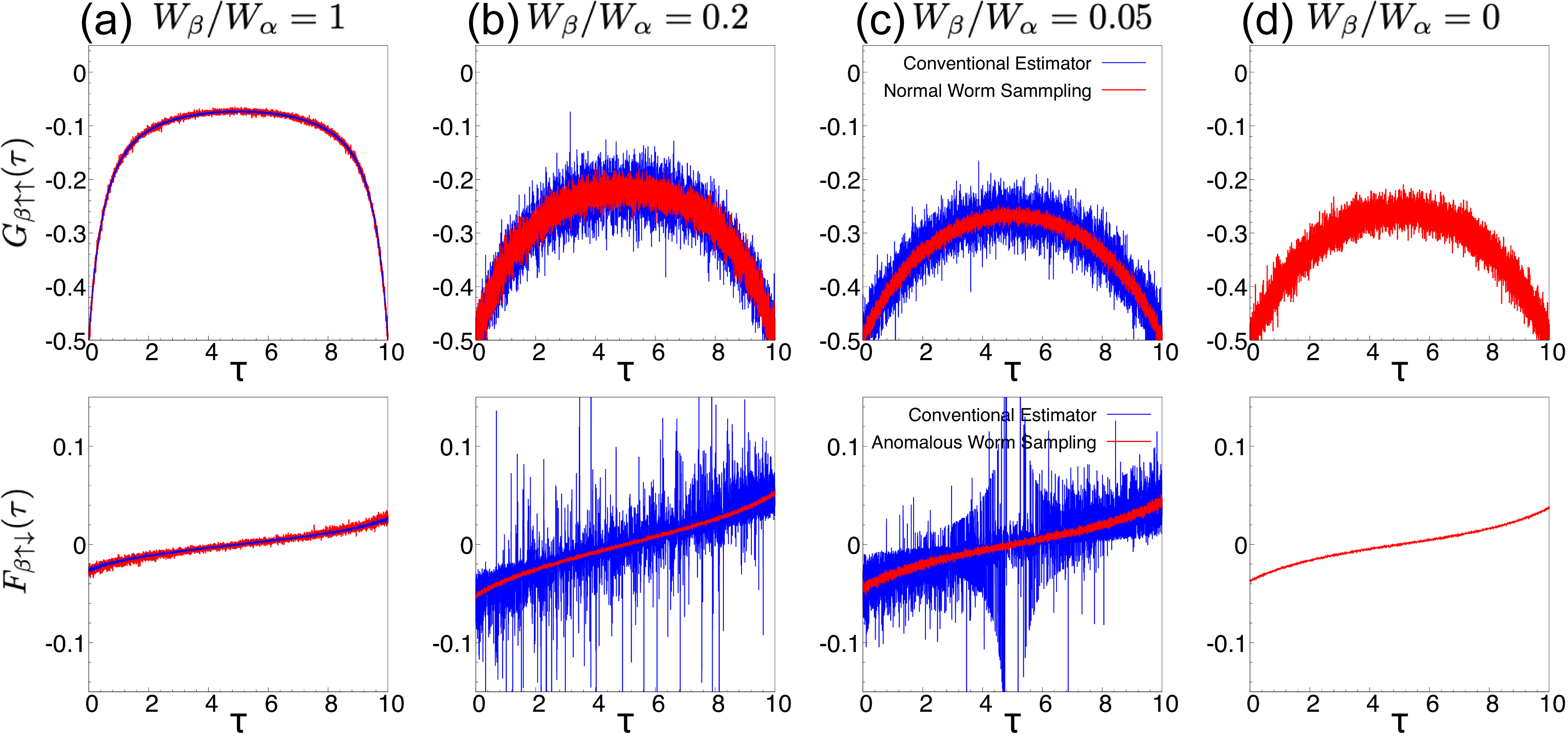}
\caption{Comparison of the normal Green's functions $G_{\beta\uparrow\uparrow}(\tau)$ 
(anomalous Green's function $F_{\beta\uparrow\downarrow}(\tau)$)
obtained using the conventional estimator and normal (anomalous) worm sampling, for the indicated band width ratios in the present system with $T=0.1$. In the limit $W_\beta=0$, the conventional measurement cannot be used. 
}
\label{fig:GF_compare}
\end{figure*}

\end{flushleft}
\end{widetext}


\begin{thebibliography}{99}
\bibitem{footnote}{In the usual proximity effect, a metal next to a SC becomes itself SC because of pairs leaking into the normal region. Here we have a SC state in the wide band which ``leaks" pairs into the narrow band though pair hopping.}
\end{thebibliography}
\end{document}